\documentclass[12pt]{iopart}

\usepackage{iopams}
\usepackage{graphicx}
\usepackage{setspace}
\usepackage{mathrsfs}
\usepackage{subfig}
\usepackage[table]{xcolor}
\usepackage{algorithm}
\usepackage{algorithmic}
\usepackage{hyperref}
\hypersetup{colorlinks, citecolor=black, filecolor=black, linkcolor=black, urlcolor=black}

\expandafter\let\csname equation*\endcsname\relax
\expandafter\let\csname endequation*\endcsname\relax

\usepackage{mathtools}
\usepackage{amsmath}

\usepackage{adjustbox}

\definecolor{grayish}{RGB}{230,230,230}

\newcommand{\refEq}[1] {(\ref{#1})}
\newcommand{\superscript}[1]{\ensuremath{^{\textrm{#1}}}}

\newcommand{\Exp}[1]{\ensuremath{\exp \left( #1 \right)}}
\newcommand{\Ln}[1]{\ensuremath{\ln \left( #1 \right)}}
\newcommand{\Order}[1]{\ensuremath{O \left( #1 \right)}}
\newcommand{\Nabla}{\ensuremath{\vec{\nabla}}}
\newcommand{\romanNum}[1]{\uppercase\expandafter{\romannumeral#1}}

\begin{document}

\title{The effect of background flow shear on gyrokinetic turbulence in the cold ion limit}

\author{Justin Ball\superscript{1}, Stephan Brunner\superscript{1}, and Ben F. McMillan\superscript{2}}

\address{\superscript{1} Ecole Polytechnique F\'{e}d\'{e}rale de Lausanne (EPFL), Swiss Plasma Center (SPC), CH-1015 Lausanne, Switzerland}

\address{\superscript{2} Centre for Fusion, Space and Astrophysics, Department of Physics, University of Warwick, Coventry CV4 7AL, UK}

\ead{Justin.Ball@epfl.ch}

\begin{abstract}

The cold ion limit of the local gyrokinetic model is rigorously taken to produce a nonlinear system of fluid equations that includes background flow shear. No fluid closure is required. By considering a simple slab geometry with magnetic drifts, but no magnetic shear, these fluid equations reduce to the Charney-Hasegawa-Mima model in the presence of flow shear. Analytic solutions to this model are found to study the impact of $E \times B$ flow shear on the stability of a single Parallel Velocity Gradient (PVG) driven mode. Additionally, the model is used to investigate the effect of background $E \times B$ flow shear on the basic three-mode nonlinear coupling, which reveals differences between zonal and non-zonal modes. These analytic results agree with gyrokinetic simulations and can serve to benchmark the numerical implementation of flow shear and nonlinear coupling.

\end{abstract}

\pacs{52.30.-q, 52.30.Gz, 52.35.Mw, 52.35.Ra, 52.65.Tt}


\section{Introduction}

The presence of sheared flow can significantly alter the turbulence in a magnetized plasma. Sheared flow is thought to be important for tokamak experiments \cite{RitzRotShearTurbSuppression1990, BurrellFlowShearAndHmode1994, WagnerHmodeReview2007} as well as many astrophysical phenomena \cite{CanutoAstroFlowShear1998, JohnsonAstroFlowShear2005, RiegerAstroFlowShear2006, ZhangAstroFlowShear2009, SquireAstroFlowShear2014}. However, its effect is not straightforward to understand. A gradient in the flow {\it parallel} to the magnetic field is able to drive turbulence through what is called the Parallel Velocity Gradient (PVG) instability \cite{dAngeloPVG1965, CattoPVG1973}. On the other hand, shear in the flow {\it perpendicular} to the magnetic field line has been shown to stabilize turbulence and reduce its ability to transport heat and particles \cite{WaltzFlowShear1998, HahmExBshear2002}.

Perpendicular flow shear in tokamaks is particularly important as it may enable the improved confinement regime H-mode \cite{WagnerHmodeReview2007}. H-mode is a robust plasma phenomenon that improves the energy confinement time of the device by roughly a factor of two and is currently viewed as essential for the success of the ITER experiment \cite{DoyleITERconfinement2007} as well as a future power plant \cite{JardinARIESATphysicsBasis2006, SorbomARC2015}. Thus, a full understanding of perpendicular flow shear is desirable to ensure that a prospective design can properly exploit H-mode.

To this end, nonlinear gyrokinetic computer codes have recently been employed to study this problem with unprecedented realism and accuracy \cite{CassonExBshear2009, HighcockRotationBifurcation2010, BarnesFlowShear2011, McMillanExBshear2011}. However, such codes, which can be hundreds of thousands of lines of code \cite{GENEwebsite, PeetersGKW2009, CandyGYRO2003}, are challenging to write, run, and interpret. Thus, analytic solutions that hold in clearly defined parameter regimes can be very valuable. They can illuminate aspects of the underlying physics and serve as benchmarks to verify the code and its execution is correct.

Such analytic efforts, of course, predate gyrokinetic codes and the supercomputers they require. The PVG instability was originally discovered using simple fluid equations in a slab geometry without magnetic shear \cite{dAngeloPVG1965}. Subsequent work rigorously derived PVG from kinetic theory \cite{SmithPVG1968} and included the effect of magnetic shear \cite{CattoPVG1973}. A more recent analytic calculation \cite{WaelbroeckFlowShearUnderstanding1992} took the cold ion limit and calculated how PVG can be stabilized by perpendicular shear flow. References \cite{NewtonFlowShearUnderstanding2010, ColeFlowShearUnderstandingEM2013, RuivoMastersThesis2014} have used fluid equations, rigorously derived from electromagnetic gyrokinetic theory, to understand the interplay between the density gradient, temperature gradient, parallel flow shear, perpendicular flow shear, and magnetic shear. Lastly, several works \cite{SchekochihinSubcriticalTurb2012, PringleFlowShear2017} study how perpendicular flow shear enables gyrokinetic simulations to exhibit ``subcritical'' turbulence, which is a nonlinear instability that is sustained despite the fact that the modes are linearly stable.

As evidenced by all of these separate works, there are several different physical effects at play that are challenging to include simultaneously and rigorously. First, the effect of perpendicular flow shear is to take the coherent structures in the plasma turbulence and, with time, gradually shear them to higher wavenumbers. This introduces an explicit time dependence that makes the evolution of any turbulence mode more complex. Previous efforts detailed above (with the exception of reference \cite{SchekochihinSubcriticalTurb2012}) have dealt with this time dependence by assuming that the solution is an exponential with a slowly-varying growth rate. This is useful for developing an intuition for the dynamics of the system, but is fundamentally an approximation that is only valid when the mode growth rate is much larger than the perpendicular flow shearing rate. This approximation breaks down when the perpendicular flow is comparable to the growth rate, which is necessarily the case to stabilize turbulence \cite{WaltzFlowShear1998}. Second, as the modes are sheared to high wavenumbers, finite gyroradius effects become increasingly important. These are generally ignored (excepting references \cite{WaelbroeckFlowShearUnderstanding1992, SchekochihinSubcriticalTurb2012}), but are important as they can determine how the mode is ultimately stabilized.

Third, for fusion devices we are fundamentally interested in properties of the nonlinearly saturated state of turbulence, rather than characteristics about its linear growth. Studying how modes nonlinearly interact can illuminate the physics of this. Without perpendicular flow shear, the usual three-wave nonlinear coupling and three-wave instability analysis only involves time-independent equations, which lead to the usual instabilities that grow exponentially in time. But, as for the linear problem described in the previous paragraph, introducing perpendicular flow shear causes the equations to gain an explicit time-dependence. Thus, the nonlinear problem gains complicated time dependence that can exhibit transient growth \cite{NewtonFlowShearUnderstanding2010, RuivoMastersThesis2014}. Interestingly, we will see that there is a surprising correspondence between the linear and nonlinear analysis.  Only recently have nonlinear dynamics been considered analytically in the presence of sheared flows. Reference \cite{NetoMastersThesis2016} uses a fluid model (without finite gyroradius effects) to investigate the stability of zonal flows. Additionally, reference \cite{KosugaPVG2017} uses the cold ion limit to study the nonlinear dynamics streamers driven by the PVG instability.

In this work, we take the cold ion limit to explore the impact of parallel and perpendicular flow shear. This limit, though not physically motivated, is attractive because it enables simple and exact results. It was used most prominently in deriving the Charney-Hasegawa-Mima equation from fluid models \cite{HasegawaMimaEquation1978} and was quickly employed to find nonlinear analytic results \cite{HasegawaThreeWaveCoupling1979}. Since then, the cold ion limit of gyrokinetics has been taken \cite{Plunk2Dturb2010, ZhuColdIonLimitGyro2010} in slab geometry and has potential applications ranging from the scrape-off layer of tokamak plasmas \cite{JorgeColdIonGBS2017} to the solar wind \cite{HowesColdIonSolarWind2009}.

In section \ref{sec:modelDeriv}, we start from the full nonlinear local gyrokinetic equation in general geometry and use the cold ion limit to rigorously derive a simple fluid model that exactly governs its behavior. No fluid closure is needed because the cold ion limit naturally provides it. From this point onwards, we consider a simple slab-like geometry without magnetic shear and show that our fluid equations reduce to the Charney-Hasegawa-Mima model \cite{HasegawaMimaEquation1978} in the presence of background flow. Ignoring magnetic shear reduces the realism of the model in space, but enables investigations of the full time evolution of a single physical Fourier mode driven by PVG and stabilized by perpendicular flow. This is done in section \ref{sec:linearResults} using a Fourier representation that is very similar to that used in gyrokinetic codes, making comparison relatively simple. Next, in section \ref{sec:nonlinearResults}, we focus on the nonlinear coupling term and study how three-wave coupling is affected by perpendicular flow shear. The results of both sections \ref{sec:linearResults} and \ref{sec:nonlinearResults} are compared against gyrokinetic simulations as they are presented. Section \ref{sec:conclusions} provides some concluding remarks.

\section{Derivation of the model}
\label{sec:modelDeriv}

We begin with the electrostatic, collisionless, local (flux-tube) $\delta f$ gyrokinetic model in a general geometry \cite{FriemanNonlinearGyrokinetics1982, ParraUpDownSym2011, AbelGyrokineticsDeriv2012}. The electrons are assumed to respond adiabatically to the motion of a single ion species. We will set the background flow in the center of our domain to be zero, but allow for the flow to vary in the $\Nabla x$ direction, which is perpendicular to the direction of the magnetic field $\hat{b}$. We refer to the gradient of the component of the flow parallel to the magnetic field line as ``parallel flow shear,'' while the gradient in the component perpendicular to the magnetic field is called ``perpendicular flow shear.'' Including both of these effects enables study of the PVG instability as well as stabilization by $\vec{E} \times \vec{B}$ flow shear. All other gradients of the plasma quantities are also assumed to vary in only in the $\Nabla x$ direction. The binormal spatial coordinate $y$ is defined such that $\Nabla y \times \Nabla x$ is in the $\hat{b}$ direction. We will denote the real-space coordinate along the field line as $z$. Note that this coordinate system is not orthogonal in general and permits the Jacobian to be different from one. However, we will only encounter $J \equiv | ( \Nabla y \times \Nabla x ) \cdot \hat{b}|^{-1}$, which is closely related to the coordinate system Jacobian. Thus, it is appropriate for both a slab representation \cite{NewtonFlowShearUnderstanding2010} (i.e. a Cartesian $y$, $x$, and $z$ such that $J = 1$) or a toroidal ballooning representation \cite{ParraUpDownSym2011} (i.e. $y$ as the binormal angle $\alpha$, $x$ as the poloidal flux $\psi$, and $z$ as the poloidal angle $\theta$ such that $J = B^{-1}$). Note that we have defined the right-handed coordinate system as $(y, x, z)$ in order to be consistent with the ballooning representation.

In this context, the real-space ion gyrokinetic equation is given by
\begin{align}
  \left( \frac{\partial}{\partial t} - \omega_{V \perp} x \frac{\partial}{\partial y} \right) & \left( \overline{h}_{i} - \frac{Z_{i} e F_{Mi}}{T_{i}} \left\langle \overline{\phi} \right\rangle \right) + v_{||} \hat{b} \cdot \Nabla \overline{h}_{i} + a_{||i} \frac{\partial \overline{h}_{i}}{\partial v_{||}} + \frac{1}{B} \left( \Nabla \langle \overline{\phi} \rangle \times \Nabla \overline{h}_{i} \right) \cdot \hat{b} \nonumber \\ 
  &+ \vec{v}_{Mi} \cdot \Nabla \overline{h}_{i} = \frac{1}{B} \frac{\partial \left\langle \overline{\phi} \right\rangle}{\partial y} F_{Mi} \left[ \frac{1}{l_{n}} + \left( \frac{m_{i} v^{2}}{2 T_{i}} - \frac{3}{2} \right) \frac{1}{l_{T}} + \frac{m_{i} v_{||}}{T_{i}} \omega_{V||} \right] ,  \label{eq:realspaceGKeq}
\end{align}
where $t$ is the time coordinate and the velocity-space coordinates are the velocity parallel to the magnetic field $v_{||}$ and the magnetic moment $\mu \equiv v_{\perp}^{2}/(2 B)$ (which is defined with the perpendicular velocity $v_{\perp}$). The magnetic drifts are given by
\begin{align}
  \vec{v}_{Mi} \equiv \frac{v_{||}^{2}}{\Omega_{i}} \hat{b} \times \vec{\kappa} + \frac{\mu}{\Omega_{i}} \hat{b} \times \Nabla B
\end{align}
and the linear parallel acceleration (i.e. the mirror effect) is
\begin{align}
  a_{||i} \equiv - \mu \hat{b} \cdot \Nabla B ,
\end{align}
where $\vec{\kappa} \equiv \hat{b} \cdot \Nabla \hat{b}$ is the magnetic field curvature vector. The background gradients are the shear in the perpendicular flow
\begin{align}
  \omega_{V \perp} \equiv - \frac{d}{dx} \left( \frac{\vec{E}_{0} \times \hat{b}}{B} \cdot \Nabla y \right) = \frac{\partial}{\partial x} \left( \frac{1}{J B} \frac{d \Phi_{0}}{d x} \right) ,
\end{align}
the shear in the parallel flow $\omega_{V||}$, the ion density gradient scale length $l_{n}^{-1} \equiv - J^{-1} d \Ln{n_{i}}/dx$, and the ion temperature gradient scale length $l_{T}^{-1} \equiv - J^{-1} d \Ln{T_{i}}/dx$. The shear in the parallel flow is $\omega_{V||} \equiv - J^{-1} ( R B_{\zeta} / B ) d (V/R) / d x$ in toroidal geometry and $\omega_{V||} \equiv - J^{-1} (\hat{b} \cdot \hat{\zeta} ) d V/dx$ in slab geometry, where $V$ is the flow velocity, $R$ is the major radius, and $\hat{\zeta}$ is the direction of the flow. Additionally, $B$ is the magnitude of the magnetic field, $\vec{E}_{0} = - \Nabla \Phi_{0}$ is the background electric field, $\Phi_{0} = \Phi_{0} ( x )$ is the background electrostatic potential, $Z_{i}$ is the ion charge number, $e$ is the elementary charge, $F_{M i} \equiv n_{i} \left(m_{i} / 2 \pi T_{i} \right)^{3/2} \Exp{-m_{i} v^2 / 2 T_{i}}$ is the background ion Maxwellian distribution function, $n_{i}$ is the background ion number density, $m_{i}$ is the ion mass, $\Omega_{i} \equiv Z_{i} e B/m_{i}$ is the ion gyrofrequency, $T_{i}$ is the background ion temperature, and $v$ is the particle speed coordinate. The unknowns in this equation are $\overline{h}_{i} \equiv \overline{\delta f}_{i} + \left( Z_{i} e F_{Mi} / T_{i} \right) \overline{\phi}$, the nonadiabatic portion of the fluctuating ion distribution function in real-space, and $\overline{\phi}$, the fluctuating electrostatic potential in real-space. Note that $\overline{\delta f}_{i}$ is the fluctuating ion distribution function in real-space and $\left\langle \ldots \right\rangle \equiv (2 \pi)^{-1} \oint_{0}^{2 \pi} d \varphi \left( \ldots \right)$ is the particle gyroaverage over gyro-angle (taken at constant particle guiding center). Since equation \refEq{eq:realspaceGKeq} has two unknowns, $\overline{h}_{i}$ and $\overline{\phi}$, we also require the real-space quasineutrality equation to close the system:
\begin{align}
  Z_{i} e \left. \int \right|_{x} d^{3}v ~ \overline{h}_{i} = \left( \frac{Z_{i}^{2} e^{2} n_{i}}{T_{i}} + \frac{e^{2} n_{e}}{T_{e}} \right) \overline{\phi} , \label{eq:realspaceQNeq}
\end{align}
where the species index indicates either $i$ for ions or $e$ for electrons and $\left. \int \right|_{x} d^{3}v$ indicates that the integral must be taken at constant particle position (not the guiding center position).

Since we are working in the local flux-tube limit of $\delta f$ gyrokinetics, the background gradients are fixed constants and it is appropriate to use periodic boundary conditions in the directions perpendicular to the magnetic field lines. Thus, it is convenient to perform a Fourier analysis of $\overline{h}_{i}$ and $\overline{\phi}$ in the radial and binormal directions (e.g. $\overline{\phi} = \sum_{k_{x}} \sum_{k_{y}} \phi \Exp{i k_{x} x + i k_{y} y}$). This has the advantage of converting the averages over gyro-angle into Bessel functions \cite{ParraUpDownSym2011, SugamaNonlinearGyrokinetics1998}. The only caveat is that, because the nonlinear term includes a product of $\overline{h}_{i}$ and $\overline{\phi}$, the Fourier transform includes a convolution (which involves three modes). In Fourier-space, equation \refEq{eq:realspaceGKeq} becomes \cite{ParraUpDownSym2011, SugamaNonlinearGyrokinetics1998}
\begin{align}
  \left( \frac{\partial}{\partial t} \right. &+ \left. \omega_{V \perp} k_{y} \frac{\partial}{\partial k_{x}} \right) g_{i} + v_{||} \hat{b} \cdot \Nabla g_{i} + i \vec{k}_{\perp} \cdot \vec{v}_{Mi} g_{i} + a_{||i} \frac{\partial g_{i}}{\partial v_{||}} \label{eq:GKeq} \\
  &+ \frac{1}{B} \sum_{\vec{k}'} \left( \vec{k}' \times \vec{k}'' \right) \cdot \hat{b} \left( g_{i}' + \frac{Z_{i} e F_{Mi}}{T_{i}} \phi' J_{0}\left( k_{\perp}' \rho_{i}\right) \right) \phi'' J_{0}\left( k_{\perp}'' \rho_{i}\right) \nonumber \\
  &= - \frac{Z_{i} e F_{Mi}}{T_{i}} \left[ v_{||} \hat{b} \cdot \Nabla \left( J_{0} \left(k_{\perp} \rho_{i}\right) \phi \right) + i \vec{k}_{\perp} \cdot \vec{v}_{Mi} J_{0} \left(k_{\perp} \rho_{i}\right) \phi - a_{||i} \frac{m_{i} v_{||}}{T_{i}} J_{0} \left(k_{\perp} \rho_{i}\right) \phi \right] \nonumber \\
  &+ i \frac{k_{y}}{B} J_{0} \left(k_{\perp} \rho_{i}\right) \phi F_{Mi} \left[ \frac{1}{l_{n}} + \left( \frac{m_{i} v^{2}}{2 T_{i}} - \frac{3}{2} \right) \frac{1}{l_{T}} + \frac{m_{i} v_{||}}{T_{i}} \omega_{V||} \right] , \nonumber
\end{align}
where the summation is performed over the full $k'_{x} \in \left( - \infty, \infty \right)$, $k'_{y} \in \left( - \infty, \infty \right)$ plane, the nonlinear coupling condition of the convolution is $\vec{k}'' = \vec{k} - \vec{k}'$, and the prime and double prime symbols indicate which wavenumber is used in evaluating the quantity. Here $k_{x}$ and $k_{y}$ are the $x$ and $y$ Fourier wavenumbers, $g_{i} (\vec{k}) \equiv h_{i} (\vec{k}) - \left( Z_{i} e F_{Mi} / T_{i} \right) J_{0} \left( k \rho_{i} \right) \phi (\vec{k})$ is the complementary ion distribution function in Fourier-space, $J_{0} \left( \ldots \right)$ is the $0^{\text{th}}$ order Bessel function of the first kind, and $\rho_{i} \equiv \sqrt{2 \mu B} / \Omega_{i}$ is the ion gyroradius.

In Fourier-space, the quasineutrality equation becomes
\begin{align}
  \int d^{3}v ~ g_{i} J_{0}\left(k_{\perp} \rho_{i}\right) = \frac{e n_{i}}{T_{e}} \left\{ Z_{i} \frac{T_{e}}{T_{i}} \left[ 1 - I_{0} \left( k_{\perp}^{2} \rho_{th i}^{2} \right) \exp \left( - k_{\perp}^{2} \rho_{th i}^{2} \right) \right] + 1 \right\} \phi , \label{eq:QNeq}
\end{align}
where $\rho_{th i} \equiv \sqrt{T_{i} / m_{i}} / \Omega_{i}$ is the ion thermal gyroradius and the integral is now taken at constant guiding center position. Such a Fourier representation facilitates comparison to (and understanding of) gyrokinetic codes, most of which use such a representation.

Next, we rename the dummy variables $\vec{k}' \rightarrow \vec{k}''$ and $\vec{k}'' \rightarrow \vec{k}'$ in equation (\ref{eq:GKeq}), then sum the result with the original equation (\ref{eq:GKeq}) to make the $(\vec{k}'$, $\vec{k}'')$ symmetry of the nonlinear term explicit. By changing to the double-shearing coordinate system \cite{CooperDoubleShearing1988, HameiriDoubleShearing1990}, we can elegantly treat both the effects of flow shear and magnetic shear. This coordinate system accounts for the fact that magnetic shear causes an eddy to be radially sheared as you move along a field line, while perpendicular flow shear causes an eddy to be radially sheared as time progresses. A more detailed explanation is given in reference \cite{NewtonFlowShearUnderstanding2010}. In Fourier-space, this coordinate transform is given by
\begin{align}
  K_{x} &\equiv k_{x} - k_{y} \omega_{V\perp} t - k_{y} \frac{z}{l_{s}} \label{eq:flowShearTransform} \\
  Z &\equiv z + u_{f} t , \label{eq:magShearTransform}
\end{align}
where $u_{f} \equiv \omega_{V\perp} l_{s}$ is the velocity at which the unsheared eddy appears to move along the field line. The parameter $l_{s}$ is the global magnetic shear scale length, which is $l_{s}^{-1} \equiv d (\hat{b} \cdot \Nabla y ) / dx$ in slab coordinates or $l_{s}^{-1} \equiv d q / dx$ in ballooning coordinates (where $q$ is the safety factor). In these new coordinates, the gyrokinetic equation becomes
\begin{align}
  \left. \frac{\partial g_{i}}{\partial t} \right|_{K_{x}} &+ \left( v_{||} \hat{b} \cdot \Nabla Z + u_{f} \right) \frac{\partial g_{i}}{\partial Z} + i \vec{k}_{\perp} \cdot \vec{v}_{Mi} g_{i} + a_{||i} \frac{\partial g_{i}}{\partial v_{||}} \label{eq:GKeqSym} \\
  +& \frac{1}{2 B} \sum_{\vec{K}'} \left( \vec{K}' \times \vec{K}'' \right) \cdot \hat{b} \left( g_{i}' \phi'' J_{0}\left(k'' \rho_{i}\right) - g_{i}'' \phi' J_{0}\left(k' \rho_{i}\right) \right) \nonumber \\
  =& - \frac{Z_{i} e F_{Mi}}{T_{i}} \left[ v_{||} \hat{b} \cdot \Nabla Z \frac{\partial}{\partial Z} \left( J_{0} \left(k_{\perp} \rho_{i}\right) \phi \right) + i \vec{k}_{\perp} \cdot \vec{v}_{Mi} J_{0} \left(k_{\perp} \rho_{i}\right) \phi - a_{||i} \frac{m_{i} v_{||}}{T_{i}} J_{0} \left(k_{\perp} \rho_{i}\right) \phi \right] \nonumber \\
  +& i \frac{k_{y}}{B} J_{0} \left(k_{\perp} \rho_{i}\right) \phi F_{Mi} \left[ \frac{1}{l_{n}} + \left( \frac{m_{i} v^{2}}{2 T_{i}} - \frac{3}{2} \right) \frac{1}{l_{T}} + \frac{m_{i} v_{||}}{T_{i}} \omega_{V||} \right] , \nonumber
\end{align}
where the summation is still over the full plane and $\vec{K}'' = \vec{K} - \vec{K}'$. Note that, since the time derivative is now taken at constant $K_{x}$, the quantities of $k_{x}$ appearing in $\vec{k}_{\perp}$ and the Bessel functions have gained an explicit time dependence.

At this point, as in reference \cite{HasegawaThreeWaveCoupling1979}, we will take the cold ion limit $T_{i} \ll Z_{i} T_{e}$. Note that we maintain $k_{\perp} \rho_{S} \sim 1$, where $\rho_{S} \equiv \sqrt{Z_{i} T_{e} / m_{i}} / \Omega_{i}$ is the sound gyroradius. In this limit, $J_{0} \left( k_{\perp} \rho_{i} \right) \rightarrow 1$, while the quasineutrality condition retains the lowest order polarization drift. Thus, the gyrokinetic equation becomes
\begin{align}
  \left. \frac{\partial g_{i}}{\partial t} \right|_{K_{x}} &+ \left( v_{||} \hat{b} \cdot \Nabla Z + u_{f} \right) \frac{\partial g_{i}}{\partial Z} + i \vec{k}_{\perp} \cdot \left( \frac{v_{||}^{2}}{\Omega_{i}} \hat{b} \times \vec{\kappa} + \frac{\mu}{\Omega_{i}} \hat{b} \times \Nabla B \right) g_{i} - \mu \hat{b} \cdot \Nabla Z \frac{\partial B}{\partial Z} \frac{\partial g_{i}}{\partial v_{||}} \nonumber \\
  +& \frac{1}{2 B} \sum_{\vec{K}'} \left( \vec{K}' \times \vec{K}'' \right) \cdot \hat{b} \left( g_{i}' \phi'' - g_{i}'' \phi' \right) \label{eq:GKeqColdIon} \\
  =& - \frac{Z_{i} e F_{Mi}}{T_{i}} \left[ v_{||} \hat{b} \cdot \Nabla Z \frac{\partial \phi}{\partial Z} + i \vec{k}_{\perp} \cdot \left( \frac{v_{||}^{2}}{\Omega_{i}} \hat{b} \times \vec{\kappa} + \frac{\mu}{\Omega_{i}} \hat{b} \times \Nabla B \right) \phi + \mu \hat{b} \cdot \Nabla Z \frac{\partial B}{\partial Z} \frac{m_{i} v_{||}}{T_{i}} \phi \right] \nonumber \\
  +& i \frac{k_{y}}{B} \phi F_{Mi} \left[ \frac{1}{l_{n}} + \left( \frac{m_{i} v^{2}}{2 T_{i}} - \frac{3}{2} \right) \frac{1}{l_{T}} + \frac{m_{i} v_{||}}{T_{i}} \omega_{V||} \right] \nonumber
\end{align}
and quasineutrality becomes
\begin{align}
  \delta n \equiv \int d^{3}v ~ g_{i} = \frac{e n_{i}}{T_{e}} \left( 1 + k_{\perp}^{2} \rho_{S}^{2} \right) \phi . \label{eq:QNeqColdIons}
\end{align}
It is important to note the finite sound gyroradius effect that survives in the quasineutrality equation. This is due to the ion polarization drift and will have important consequences later in this work. Taking velocity-space moments of equation \refEq{eq:GKeqColdIon}, gives a fluid model. We find that no closure is needed because, in the cold ion limit, the right-hand side is zero for all except the density and parallel velocity moments. Thus, $\int d^{3}v ~ v_{||}^{a} \mu^{b} g_{i} = 0$ for integers $a \geq 2$ and $b \geq 1$. The density moment is
\begin{align}
  \left(\left. \frac{\partial}{\partial t} \right|_{K_{x}} + u_{f} \frac{\partial}{\partial Z} \right) \delta n &+ n_{i} \hat{b} \cdot \Nabla Z \frac{\partial \delta u_{||}}{\partial Z} + \frac{1}{2 B} \sum_{\vec{K}'} \left( \vec{K}' \times \vec{K}'' \right) \cdot \hat{b} \left(\delta n' \phi'' - \delta n'' \phi' \right) \label{eq:densityMoment} \\
  &+ i \frac{e n_{i}}{T_{e}} \left( k_{x} \rho_{S} \omega_{Mx} + k_{y} \rho_{S} \omega_{My} \right) \phi = 0 , \nonumber
\end{align}
where
\begin{align}
  \omega_{Mx} &\equiv c_{S} \left( \frac{1}{J} \frac{\partial \ln B}{\partial y} + \frac{\kappa_{y}}{J} \right) \\
  \omega_{My} &\equiv - c_{S} \left( \frac{1}{l_{n}} + \frac{1}{J} \frac{\partial \ln B}{\partial x} + \frac{\kappa_{x}}{J} \right)
\end{align}
contain the effect of the magnetic drifts and density gradient, $\kappa_{x} \equiv \partial \vec{r} / \partial x \cdot \vec{\kappa}$ and $\kappa_{y} \equiv \partial \vec{r} / \partial y \cdot \vec{\kappa}$ are the components of the magnetic field line curvature, and $c_{S} \equiv \sqrt{Z_{i} T_{e}/m_{i}}$ is the sound speed. Substituting quasineutrality (i.e. equation \refEq{eq:QNeqColdIons}) gives an equation for the evolution of $\phi$:
\begin{align}
  \left(\left. \frac{\partial}{\partial t} \right|_{K_{x}} \right. &+ \left. u_{f} \frac{\partial}{\partial Z} \right) \left[ \left( 1 + k_{\perp}^{2} \rho_{S}^{2} \right) \phi \right] + \frac{T_{e}}{e} \hat{b} \cdot \Nabla Z \frac{\partial \delta u_{||}}{\partial Z}  \label{eq:phiEvolution} \\
  &+ \frac{1}{2 B} \sum_{\vec{K}'} \left( \vec{K}' \times \vec{K}'' \right) \cdot \hat{b} \left( k_{\perp}'^{2} - k_{\perp}''^{2} \right) \rho_{S}^{2} \phi' \phi'' + i \left( k_{x} \rho_{S} \omega_{Mx} + k_{y} \rho_{S} \omega_{My} \right) \phi = 0 . \nonumber
\end{align}
Lastly, the parallel velocity moment, $\delta u_{||} \equiv n_{i}^{-1} \int d^{3}v ~ v_{||} g_{i}$, of equation \refEq{eq:GKeqColdIon} is
\begin{align}
  \left(\left. \frac{\partial}{\partial t} \right|_{K_{x}} + u_{f} \frac{\partial}{\partial Z} \right) \delta u_{||} &+ \frac{1}{2 B} \sum_{\vec{K}'} \left( \vec{K}' \times \vec{K}'' \right) \cdot \hat{b} \left(\delta u_{||}' \phi'' - \delta u_{||}'' \phi' \right) \label{eq:velocityMoment} \\
  &+ \frac{Z_{i} e}{m_{i}} \hat{b} \cdot \Nabla Z \left( \frac{\partial \phi}{\partial Z} +\frac{\partial \ln B}{\partial Z} \phi \right) - i k_{y} \frac{\omega_{V||}}{B} \phi = 0 . \nonumber
\end{align}
Note that, in the cold ion limit, the effect of the ion temperature gradient vanishes entirely from the model. In real-space, equations \refEq{eq:phiEvolution} and \refEq{eq:velocityMoment} are
\begin{align}
  \left(\left. \frac{\partial}{\partial t} \right|_{Y} \right. &+ \left. u_{f} \frac{\partial}{\partial Z} \right) \left[ \left( 1 - \rho_{S}^{2} \nabla_{\perp}^{2} \right) \phi \right] + \frac{T_{e}}{e} \hat{b} \cdot \Nabla Z \frac{\partial \delta u_{||}}{\partial Z}  \label{eq:phiEvolutionRealSpace} \\
  &+ \frac{1}{B} \left( \rho_{S}^{2} \Nabla \left( \nabla_{\perp}^{2} \phi \right) \times \Nabla \phi \right) \cdot \hat{b} + \left( \omega_{Mx} \rho_{S} \frac{\partial \phi}{\partial x} + \omega_{My} \rho_{S} \frac{\partial \phi}{\partial Y} \right) = 0 \nonumber
\end{align}
and
\begin{align}
  \left(\left. \frac{\partial}{\partial t} \right|_{Y} + u_{f} \frac{\partial}{\partial Z} \right) \delta u_{||} &+ \frac{1}{B} \left( \Nabla \phi \times \Nabla \delta u_{||} \right) \cdot \hat{b} \label{eq:velocityMomentRealSpace} \\
  &+ \frac{Z_{i} e}{m_{i}} \hat{b} \cdot \Nabla Z \left( \frac{\partial \phi}{\partial Z} +\frac{\partial \ln B}{\partial Z} \phi \right) - \frac{\omega_{V||}}{B} \frac{\partial \phi}{\partial Y} = 0 , \nonumber
\end{align}
where $\Nabla_{\perp} \equiv \Nabla - \hat{b} ( \partial/\partial z )$. Note that in real-space, the double shearing coordinate system \cite{NewtonFlowShearUnderstanding2010} is given by equation \refEq{eq:magShearTransform} and $Y \equiv y - x ( z + u_{f} t ) / l_{s}$, which means that $\partial / \partial x = \left. \partial / \partial x \right|_{Y} - (Z/l_{s}) \partial / \partial Y$.

Equations \refEq{eq:phiEvolution} and \refEq{eq:velocityMoment} form a closed system that governs the electrostatic, collisionless local gyrokinetic model in the cold ion limit. This fluid model is tremendously simpler than the six-dimensional, integro-differential system of gyrokinetics. Yet, as long as the limit of $T_{i} \ll Z_{i} T_{e}$ is satisfied, the two are equivalent.

This cold ion model is simple enough to enable analytic results for the {\it linear} dynamics in a slab with magnetic shear, as was done by Waelbroeck, {\it et al.} \cite{WaelbroeckFlowShearUnderstanding1992}. However, in this work we will sacrifice realism in space in order to enable a more realistic treatment in time. Instead of a slab geometry with magnetic shear, we will use a slab {\it without} magnetic shear (but maintain simple magnetic drifts). Accordingly, instead of having to solve differential equations in space, as was done by Waelbroeck, we will solve them in time. This will permit an investigation of the full time evolution of modes under the effect of finite $\omega_{V\perp}$ as well as their nonlinear interaction.

Without magnetic shear, we can let $l_{s} \rightarrow \infty$. Moreover, to prevent the coordinate system from diverging, we must also set $u_{f} = 0$. This is physically motivated because, without magnetic shear, the unsheared eddy no longer has any apparent motion along the field line. Thus, we can ignore the last term in equations \refEq{eq:flowShearTransform} and \refEq{eq:magShearTransform} as well as the terms containing $u_{f}$ in equations \refEq{eq:phiEvolution} and \refEq{eq:velocityMoment}. Importantly, ignoring magnetic shear allows us to apply a standard periodic boundary condition in the parallel direction, rather than the more complex twist-and-shift boundary condition \cite{BeerBallooingCoordinates1995}. However, this simplification is also a limitation as it eliminates mode coupling through the parallel boundary condition. Such coupling occurs in toroidal devices, but will not be included in our model. 
Lastly, we will assume that $dB/dZ = 0$, so that $\hat{b}$ becomes a direction of symmetry. We will maintain magnetic drifts, but also assume that they are constant in $Z$. This is applicable to geometries like purely toroidal field lines or straight field lines with a perpendicular gradient in the field strength. With these simplifications, it becomes useful for notational simplicity to adopt a Cartesian orthonormal coordinate system such that $| \Nabla x | = | \Nabla y | = | \Nabla Z | = 1$, $\hat{b} \cdot \Nabla Z = 1$, and all the metric coefficients are $1$. Additionally, we can Fourier analyze in the parallel direction and replace $d\phi/dZ$ with a parallel wavenumber $i k_{||} \phi$. Doing so is only possible because all equilibrium quantities have become independent of $Z$. In this geometry, our model becomes
\begin{align}
  \left. \frac{\partial}{\partial t} \right|_{K_{x}} & \left[ \left( 1 + k_{\perp}^{2} \rho_{S}^{2} \right) \phi \right] + i k_{||} \frac{T_{e}}{e} \delta u_{||} + \frac{1}{2 B} \sum_{\vec{K}'} \left( \vec{K}' \times \vec{K}'' \right) \cdot \hat{b} \left( k_{\perp}'^{2} - k_{\perp}''^{2} \right) \rho_{S}^{2} \phi' \phi'' \label{eq:phiEvolutionFinal} \\
  &+ i \left( k_{x} \rho_{S} \omega_{Mx} + k_{y} \rho_{S} \omega_{My} \right) \phi = 0 \nonumber
\end{align}
and
\begin{align}
  \left. \frac{\partial}{\partial t} \right|_{K_{x}} \delta u_{||} &+ \frac{1}{2 B} \sum_{\vec{K}'} \left( \vec{K}' \times \vec{K}'' \right) \cdot \hat{b} \left(\delta u_{||}' \phi'' - \delta u_{||}'' \phi' \right) = - i \left( \frac{Z_{i} e}{m_{i}} k_{||} - k_{y} \frac{\omega_{V||}}{B} \right) \phi , \label{eq:velocityMomentFinal}
\end{align}
which is equivalent to the Charney-Hasegawa-Mima model \cite{HasegawaMimaEquation1978} with the addition of background flow. We note that in the slab limit of most toroidal gyrokinetic codes $\omega_{Mx} = 0$ and $\omega_{My} = -c_{S}/l_{n}$.

\section{Single mode slab results}
\label{sec:linearResults}

By restricting our analysis to the evolution of just a single mode, the nonlinear terms vanish from the $\phi$ evolution equation and the parallel velocity moment (i.e. equations \refEq{eq:phiEvolutionFinal} and \refEq{eq:velocityMomentFinal}). Qualitatively similar systems have been analyzed in the past. Some used {\it computational} approaches to investigate realistic geometries \cite{WaltzFlowShear1998, CooperDoubleShearing1988}, while others obtained analytic results for a simplified fluid model \cite{PringleFlowShear2017}. The physics discussed in these studies are useful in interpreting the results of this section, which will use the cold ion limit and above simplifications to enable rigorous {\it analytic} results.

Taking the time derivative of the equation \refEq{eq:phiEvolutionFinal} and substituting equation \refEq{eq:velocityMomentFinal} as well as the ansatz
\begin{align}
  \phi = \hat{\phi} \left( 1 + k_{\perp}^{2} \rho_{S}^{2} \right)^{-1} \Exp{- i \left. \int \right|_{K_{x}} dt ~ \frac{k_{x} \rho_{S} \omega_{Mx} + k_{y} \rho_{S} \omega_{My}}{1 + k_{\perp}^{2} \rho_{S}^{2}}} \label{eq:hatDef}
\end{align}
produces
\begin{align}
  \left( 1 + k_{\perp}^{2} \rho_{S}^{2} \right) \left. \frac{\partial^{2} \hat{\phi}}{\partial t^{2}} \right|_{K_{x}} &- i \left( k_{x} \rho_{S} \omega_{Mx} + k_{y} \rho_{S} \omega_{My} \right)  \left. \frac{\partial \hat{\phi}}{\partial t} \right|_{K_{x}} - \left( k_{||} c_{S} k_{y} \rho_{S} \omega_{V||} - k_{||}^{2} c_{S}^{2} \right) \hat{\phi} = 0 . \label{eq:phiLinearEvolution}
\end{align}
This ordinary differential equation determines the full time evolution of a single mode in the presence of parallel and perpendicular velocity shear. While it appears simple, due to the perpendicular flow shear dependence of $k_{x} = K_{x} + k_{y} \omega_{V\perp}t$ (see equation \refEq{eq:flowShearTransform}), $k_{\perp}^{2}$ is quadratic in time and $k_{x}$ is linear. Additionally, we note that the ansatz of equation \refEq{eq:hatDef} is important to keep in mind because, even with $\omega_{V||} = k_{||} = 0$ the mode will still evolve due to the variation of the $1 + k_{\perp}^{2} \rho_{S}^{2}$ factor as it is advected by perpendicular flow shear.

While equation \refEq{eq:phiLinearEvolution} is complicated, it still has an analytic solution given by
\begin{align}
  \hat{\phi} = C_{\alpha} \prescript{}{2}{F}_{1} \left( \tilde{a}, \tilde{b}; \tilde{c}; \tilde{t} \right) + C_{\beta} \tilde{t}^{1-\tilde{c}} \prescript{}{2}{F}_{1} \left( \tilde{a}+1-\tilde{c}, \tilde{b}+1-\tilde{c}; 2-\tilde{c}; \tilde{t} \right) , \label{eq:phiLinearSol}
\end{align}
where $\prescript{}{2}{F}_{1}$ is the Gaussian hypergeometric function \cite{ArfkenMath2013},
\begin{align}
  \tilde{t} &\equiv \frac{2 a_{2} t + a_{1} + \sqrt{a_{1}^{2} - 4 a_{0} a_{2}}}{2 \sqrt{a_{1}^{2} - 4 a_{0} a_{2}}} \\
  \tilde{a} &\equiv \frac{b_{1} - a_{2} + \sqrt{ \left( b_{1} - a_{2} \right)^{2} - 4 a_{2} c_{0} }}{2 a_{2}} \\
  \tilde{b} &\equiv \frac{c_{0}}{a_{2}} \frac{1}{a} \\
  \tilde{c} &\equiv \frac{-2 a_{2} b_{0} + b_{1} \left( a_{1} + \sqrt{a_{1}^{2} - 4 a_{0} a_{2}} \right)}{2 a_{2} \sqrt{a_{1}^{2} - 4 a_{0} a_{2}}} ,
\end{align}
and
\begin{align}
  a_{0} &\equiv 1 + K_{x}^{2} \rho_{S}^{2} + k_{y}^{2} \rho_{S}^{2} \\
  a_{1} &\equiv 2 K_{x} k_{y} \rho_{S}^{2} \omega_{V\perp} \\
  a_{2} &\equiv k_{y}^{2} \rho_{S}^{2} \omega_{V\perp}^{2} \\
  b_{0} &\equiv -i \left( K_{x} \rho_{S} \omega_{Mx} + k_{y} \rho_{S} \omega_{My} \right) \\
  b_{1} &\equiv -i k_{y} \rho_{S} \omega_{V\perp} \omega_{Mx} \\
  c_{0} &\equiv - \left( k_{||} c_{S} k_{y} \rho_{S} \omega_{V||} - k_{||}^{2} c_{S}^{2} \right)
\end{align}
are the coefficients of the polynomials appearing in equation \refEq{eq:phiLinearEvolution}. Here $C_{j}$ for any Greek letter $j$ is an integration constant that can be calculated from the initial conditions. This analytic solution has been verified against the numerical solution of equation \refEq{eq:phiLinearEvolution} as shown in figure \ref{fig:stableButUnbounded}. Additionally, it has been verified by comparison with the local version of the gyrokinetic code GENE \cite{JenkoGENE2000, GoerlerGENE2011} as shown in figure \ref{fig:linearSolGENE}. GENE is one of the most commonly used gyrokinetic codes with the capacity to solve the full nonlinear gyrokinetic system of equations in multiple different geometries. Among many capabilities, it can model collisions, electromagnetic fluctuations, and global effects, although these features are not used in this work.

All GENE simulations in this work use $Z_{i} = 1$ and an ion temperature of $T_{i} = 10^{-4} T_{e}$ to ensure that the cold ion limit is well satisfied. To be consistent with the analytic model, all simulations also set the magnetic shear equal to zero and use the standard slab geometry in GENE. Because this slab model does not include magnetic drifts, $\omega_{Mx}$ is forced to be zero in all simulations, but $\omega_{My}$ is varied using the density gradient. Additionally, for all comparisons it was necessary to set the flux surface averaged value of the electrostatic potential to zero when GENE calculates the adiabatic electron response. This is because, without magnetic shear, the standard parallel boundary condition used by most local gyrokinetic codes causes every field line to close on itself, so flux surfaces are not formed \cite{BeerBallooingCoordinates1995}. {All simulations use the same resolution of 32 grid points in the parallel direction, 32 parallel velocity grid points, and 24 magnetic moment grid points. To ensure that the ``wavevector-remap'' scheme converges to a fairly smooth and continuous mode evolution, 128 radial wavenumbers were used \cite{McmillanRemap2017}. Since we will only ever initialize a limited number of individual modes to be finite, the number of binormal wavenumbers does not have an effect, so very low values between 2 and 10 were used.

\begin{figure}
  \centering
  \includegraphics[width=\textwidth]{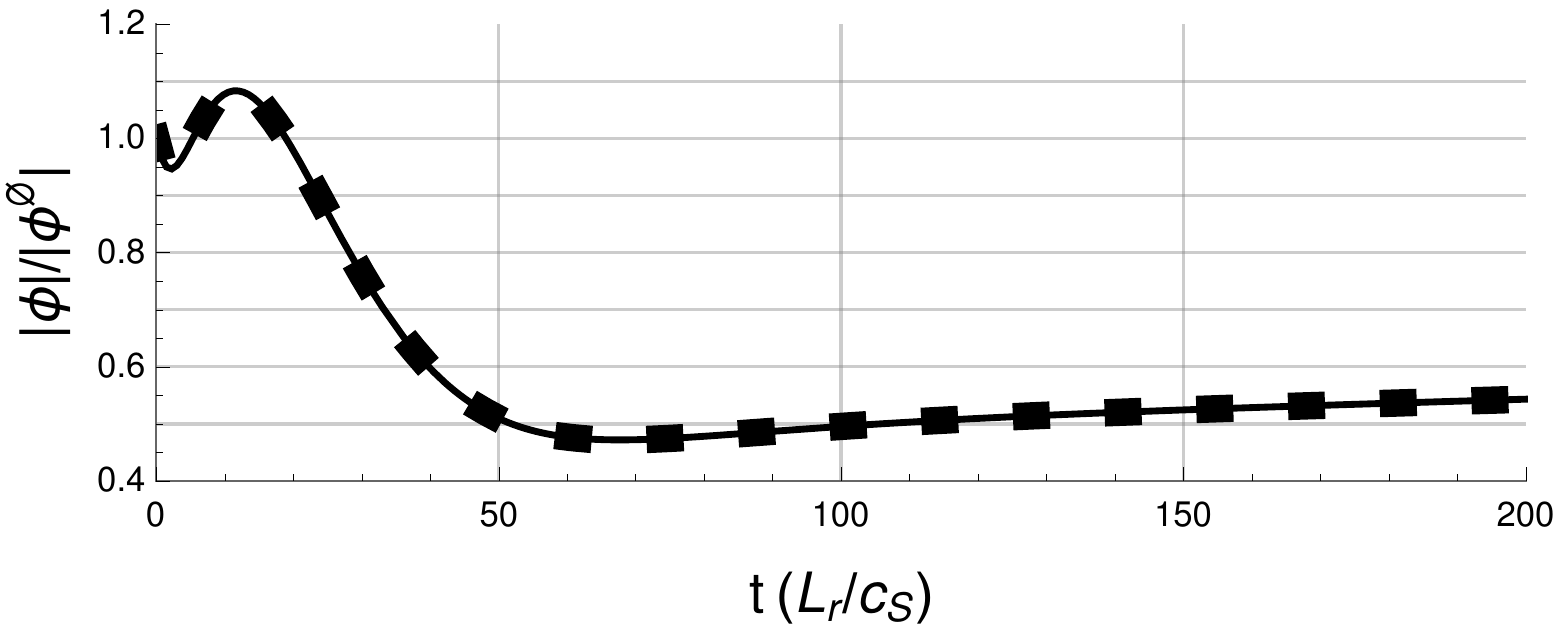}
  \caption{The numerical solution to equation \refEq{eq:phiLinearEvolution} (thin solid) and the analytic solution given by equation \refEq{eq:phiLinearSol} (thick dotted) for a PVG single mode that is advected by perpendicular flow shear. The parameters used are $\omega_{V\perp} = 0.2 c_{S}/L_{r}$, $\omega_{V||} = 1.75 c_{S}/L_{r}$, $\omega_{Mx} = 0.5 c_{S}/L_{r}$, $\omega_{My} = 0.9 c_{S}/L_{r}$, $K_{x} \rho_{S} = 1$, $k_{y} \rho_{S} = 0.3$, and $k_{||} c_{S} = k_{y} \rho_{S} \omega_{V||} / 2$, where $L_{r}$ is an arbitrary reference length.}
  \label{fig:stableButUnbounded}
\end{figure}

\begin{figure}
  \centering
  \includegraphics[width=0.6\textwidth]{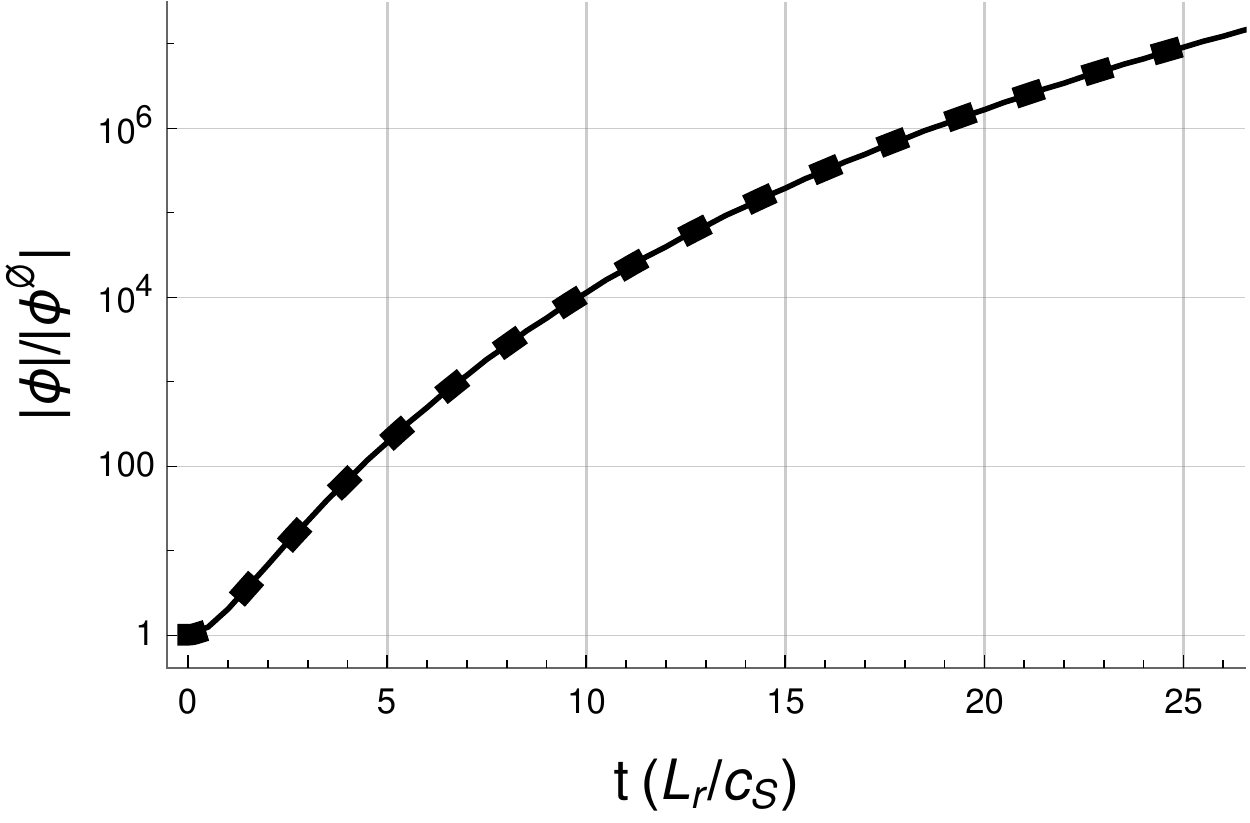}
  \caption{The mode evolution as calculated by GENE (thin solid) and the analytic solution given by equation \refEq{eq:phiLinearSol} (thick dotted) for a PVG single mode that is advected by perpendicular flow shear. The parameters used are $\omega_{V\perp} = 0.5 c_{S}/L_{r}$, $\omega_{V||} = 10 c_{S}/L_{r}$, $\omega_{Mx} = 0$, $\omega_{My} = - c_{S}/L_{r}$, $K_{x} \rho_{S} = 0$, $k_{y} \rho_{S} = 0.3$, and $k_{||} L_{r} = 1$.}
  \label{fig:linearSolGENE}
\end{figure}

Unfortunately, hypergeometric functions do not provide much physical insight. Thus, as was done in reference \cite{SchekochihinSubcriticalTurb2012}, we will find further analytic results by investigating equation \refEq{eq:phiLinearEvolution} in certain limits. We will focus on $|\omega_{V\perp} t| \ll 1$ and $|\omega_{V\perp} t| \gg 1$. The first limit can be used to understand the effect of weak perpendicular flow shear or the initial effect of arbitrary flow shear. The second limit can be used to understand the behavior of a mode after a long time as it is advected to large $k_{x}$ or the behavior of a mode with large $k_{x}$ as it is advected towards $k_{x} = 0$. One could also find solutions in the weak perpendicular flow shear limit of $\omega_{V\perp} \ll \gamma$, where $\gamma$ is the mode growth rate. We will see that this is closely related to the $|\omega_{V\perp} t| \ll 1$ limit.

\subsection{The $|\omega_{V\perp} t| \ll 1$ limit}
\label{subsec:PVGlowFlowLimit}

By looking in the limit of short time, we can obtain analytic results for situations without perpendicular flow shear as well as the early time behavior of situations with flow shear. To lowest order (i.e. $\omega_{V \perp} = 0$), we see that $k_{x} = K_{x} = \text{const}$ and $k_{\perp} = K_{x}^{2} + k_{y}^{2} = \text{const}$, so the time dependence of the wavenumber coefficients vanish. Thus, equation \refEq{eq:phiLinearEvolution} becomes a simple 2\textsuperscript{nd} order ordinary differential equation with constant coefficients. This is solved by an exponential, which can be substituted into equation \refEq{eq:hatDef} to find the full solution (without the hat on $\phi$) to be
\begin{align}
  \phi_{0} = C_{\gamma} \Exp{\left(i \omega + \gamma \right) t} + C_{\delta} \Exp{\left(i \omega - \gamma \right) t} , \label{eq:noFlowSol}
\end{align}
where the numerical subscript represents the quantity's order in the $|\omega_{V\perp} t|$ expansion and
\begin{align}
  \omega &\equiv - \frac{1}{2} \frac{K_{x} \rho_{S} \omega_{Mx} + k_{y} \rho_{S} \omega_{My}}{1 + k_{\perp}^{\varnothing 2} \rho_{S}^{2}} \label{eq:noFlowFreq} \\
  \gamma &\equiv \sqrt{\frac{k_{||} c_{S} k_{y} \rho_{S} \omega_{V||} - k_{||}^{2} c_{S}^{2}}{1 + k_{\perp}^{\varnothing 2} \rho_{S}^{2}} - \frac{1}{4} \left( \frac{K_{x} \rho_{S} \omega_{Mx} + k_{y} \rho_{S} \omega_{My}}{1 + k_{\perp}^{\varnothing 2} \rho_{S}^{2}} \right)^{2}} \label{eq:noFlowGrowthRate}
\end{align}
are the mode frequency and growth rate respectively. Here the superscript $\varnothing$ indicates the initial condition of the quantity (e.g. $k_{x}^{\varnothing} \equiv k_{x} (t=0) = K_{x}$, $k_{\perp}^{\varnothing 2} = K_{x}^{2} + k_{y}^{2}$). When the mode growth rate is much greater than the perpendicular flow shear (i.e. $\omega_{V\perp} \ll \gamma$), this solution actually holds for all time. We can simply replace $K_{x} \rightarrow k_{x}$ and $k_{\perp}^{\varnothing} \rightarrow k_{\perp}$ to capture their time dependence because they vary slowly in the $\omega_{V\perp} \ll \gamma$ expansion. Note that, when the magnetic drifts and finite sound gyroradius effects are ignored, equation \refEq{eq:noFlowGrowthRate} reduces to the typical PVG growth rate \cite{CattoPVG1973, NewtonFlowShearUnderstanding2010}.

For instability, the growth rate $\gamma$ must be real, giving the following condition on the parallel velocity gradient:
\begin{align}
  \omega_{V||} > \frac{1}{k_{||} c_{S} k_{y} \rho_{S} } \left( k_{||}^{2} c_{S}^{2} + \frac{1}{4} \frac{\left( K_{x} \rho_{S} \omega_{Mx} + k_{y} \rho_{S} \omega_{My} \right)^{2}}{1 + k_{\perp}^{\varnothing 2} \rho_{S}^{2}} \right) . \label{eq:PVGstabilityShortTime}
\end{align}
This shows that the only possible instability in this limit is PVG. As is intuitive, in the cold ion limit neither ion temperature gradient (ITG) nor drift wave instabilities can exist. Without parallel flow shear, equation \refEq{eq:noFlowSol} simply governs stable drift-sound waves as they oscillate. On the other hand, we see that a sufficiently large parallel velocity gradient will always overcome the damping and lead to instability, so long as $k_{||}$ and $k_{y}$ are finite. To maximize the growth rate, a mode should have a parallel wavenumber of $k_{||} c_{S} = k_{y} \rho_{S} \omega_{V||} / 2$, and as large of a $k_{y}$ as possible. Given this value of $k_{||}$, the stability criterion becomes
\begin{align}
  \omega_{V||}^{2} > \frac{1}{k_{y}^{2} \rho_{S}^{2}} \frac{\left( K_{x} \rho_{S} \omega_{Mx} + k_{y} \rho_{S} \omega_{My} \right)^{2}}{1 + k_{\perp}^{\varnothing 2} \rho_{S}^{2}} . \label{eq:PVGstabilityShortTimekPar}
\end{align}
Note that, due to finite sound gyroradius effects in equation \refEq{eq:noFlowGrowthRate}, the growth rate has a maximum possible value of $\gamma = |\omega_{V||}|/2$, rather than increasing without bound when $k_{y} \rightarrow \infty$ as is the usual result \cite{CattoPVG1973, NewtonFlowShearUnderstanding2010}.

The dependence of the instability growth rate on $K_{x}$ is complex as shown in figure \ref{fig:kxLinearSpectrum}. For example, when the magnetic drifts and density gradient are zero such that $\omega_{My} = 0$, it is best for $K_{x} \rho_{S} = 0$ to minimize the finite gyroradius damping. However, when $\omega_{My} \neq \omega_{Mx} = 0$ the maximum growth rate occurs at $K_{x} \rho_{S} = \pm \sqrt{2 \omega_{My}^2 - (1 + k_{y}^{2} \rho_{S}^2) \omega_{V||}^2}/\omega_{V||}$. As can be seen from equation \refEq{eq:noFlowGrowthRate}, when both $\omega_{Mx}$ and $\omega_{My}$ are non-zero the growth rate spectrum even becomes asymmetric about $K_{x} \rho_{S} = 0$.

\begin{figure}
  \centering
  \includegraphics[width=0.6\textwidth]{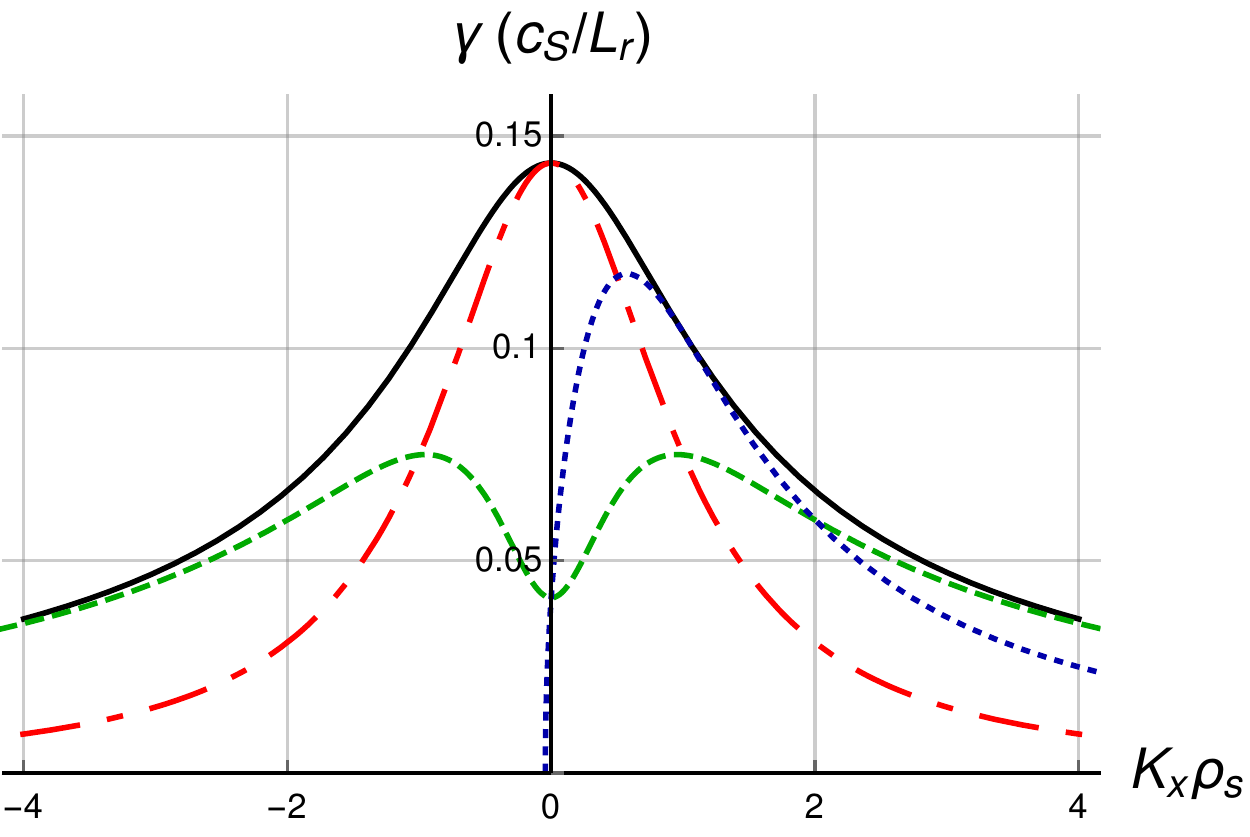}
  \caption{The mode growth rate as calculated by equation \refEq{eq:noFlowGrowthRate} for $\omega_{Mx} = \omega_{My} = 0$ (solid black), $\omega_{Mx} = -0.3 c_{S}/L_{r}$ and $\omega_{My} = 0$ (dash-dotted red), $\omega_{Mx} = 0$ and $\omega_{My} = c_{S}/L_{r}$ (dashed green), and $\omega_{Mx} = -0.3 c_{S}/L_{r}$ and $\omega_{My} = c_{S}/L_{r}$ (dotted blue). The other parameters were set to $\omega_{V\perp} = 0$, $\omega_{V||} = c_{S}/L_{r}$, $k_{y} \rho_{S} = 0.3$, and $k_{||} c_{S} = k_{y} \rho_{S} \omega_{V||} / 2$.}
  \label{fig:kxLinearSpectrum}
\end{figure}

If the magnetic drifts, density gradient, and finite gyroradius effects are neglected, these results match equation (30) of reference \cite{NewtonFlowShearUnderstanding2010} and equation (22) of reference \cite{CattoPVG1973} when $T_{i}=0$. This is true even though magnetic shear is included in the analysis of reference \cite{NewtonFlowShearUnderstanding2010}. Additionally, the effect of finite gyroradius effects is in agreement with references \cite{WaelbroeckFlowShearUnderstanding1992, KosugaPVG2017}.

To next order in $|\omega_{V\perp} t| \ll 1$, the differential equation for the mode becomes
\begin{align}
  \left( 1 + k_{\perp}^{\varnothing 2} \rho_{S}^{2} \right) \left. \frac{\partial^{2} \hat{\phi}_{1}}{\partial t^{2}} \right|_{K_{x}} &- i \left( K_{x} \rho_{S} \omega_{Mx} + k_{y} \rho_{S} \omega_{My} \right)  \left. \frac{\partial \hat{\phi}_{1}}{\partial t} \right|_{K_{x}} - \left( k_{||} c_{S} k_{y} \rho_{S} \omega_{V||} - k_{||}^{2} c_{S}^{2} \right) \hat{\phi}_{1} \nonumber \\
  &= - \left( k_{\perp}^{2} \rho_{S}^{2} \right)_{1} \left. \frac{\partial^{2} \hat{\phi}_{0}}{\partial t^{2}} \right|_{K_{x}} + i \left( k_{x} \rho_{S} \right)_{1} \omega_{Mx}  \left. \frac{\partial \hat{\phi}_{0}}{\partial t} \right|_{K_{x}} , \label{eq:linearEvolutionNextOrder}
\end{align}
Perpendicular flow shear appears only in the inhomogeneous terms through $\left( k_{\perp}^{2} \rho_{S}^{2} \right)_{1} = 2 K_{x} k_{y} \rho_{S}^{2} \omega_{V\perp} t$ and $\left( k_{x} \rho_{S} \right)_{1} = k_{y} \rho_{S} \omega_{V\perp} t$. This equation can be solved analytically. The solution to the homogeneous equation has the same form as the lowest order solution of equation \refEq{eq:noFlowSol}, while a particular solution to the inhomogeneous equation is
\begin{align}
  \hat{\phi}_{1} &= \frac{1}{4 \gamma^{2}} \Big\{ C_{\gamma} \Exp{\left( -i \omega + \gamma \right) t} \left( 1 - \gamma t \right) \left[ \left( - i \omega + \gamma \right)^{2} \left( k_{\perp}^{2} \rho_{S}^{2} \right)_{1} - i \left( -i \omega + \gamma \right) \left( k_{x} \rho_{S} \right)_{1}  \omega_{Mx} \right] \nonumber \\
  &+ C_{\delta} \Exp{\left( -i \omega - \gamma \right) t} \left( 1 + \gamma t \right) \left[ \left( - i \omega - \gamma \right)^{2} \left( k_{\perp}^{2} \rho_{S}^{2} \right)_{1} - i \left( -i \omega - \gamma \right) \left( k_{x} \rho_{S} \right)_{1}  \omega_{Mx} \right] \Big\} , \label{eq:linearSolNextOrder}
\end{align}
where we note the hat on $\hat{\phi}_{1}$. Thus, we see that the dominant effect of weak flow shear (or alternatively the first effect of flow shear to appear) is a quadratic correction to the lowest order exponential behavior. The lowest and next order analytic solutions are compared to the exact numerical solution of equation \refEq{eq:phiLinearEvolution} in figure \ref{fig:linearResultsShortTime}. Note that care must be taken in converting from $\hat{\phi}_{1}$  to $\phi_{1}$ because equation \refEq{eq:hatDef} must be expanded to next order in $|\omega_{V\perp} t| \ll 1$.

\begin{figure}
  \centering
  \includegraphics[width=0.6\textwidth]{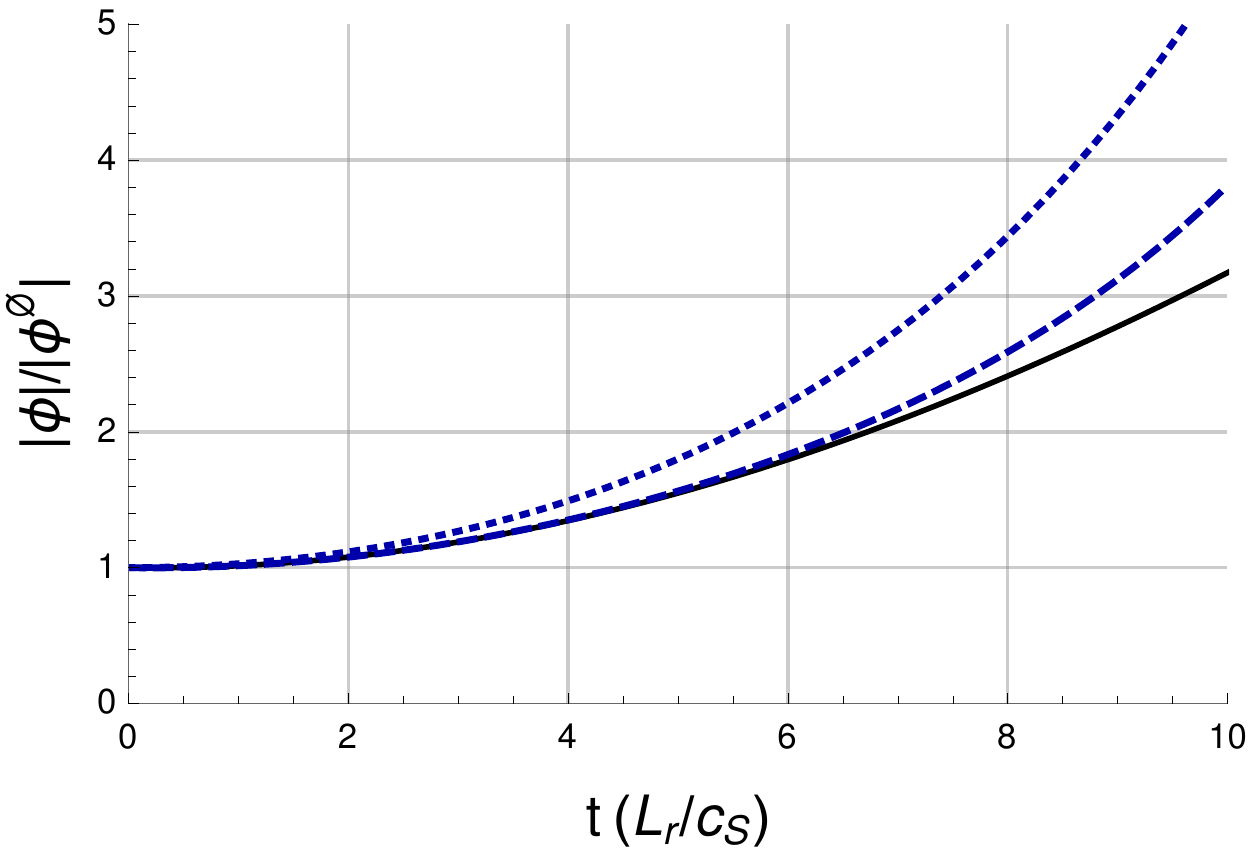}
  \caption{The mode evolution for short times as calculated by the numerical solution to equation \refEq{eq:phiLinearEvolution} (solid black), the lowest order analytic solution of equation \refEq{eq:noFlowSol} (dotted blue), or the next order solution that includes equation \refEq{eq:linearSolNextOrder} (dashed blue). The parameters used are $\omega_{V\perp} = 0.2 c_{S}/L_{r}$, $\omega_{V||} = 1.67 c_{S}/L_{r}$, $\omega_{Mx} = 0.5 c_{S}/L_{r}$, $\omega_{My} = 0$, $K_{x} \rho_{S} = 0.1$, $k_{y} \rho_{S} = 0.3$, and $k_{||} c_{S} = k_{y} \rho_{S} \omega_{V||} / 2$.}
  \label{fig:linearResultsShortTime}
\end{figure}

While equation \refEq{eq:linearSolNextOrder} is fairly complicated, it can be further simplified to give an expected result about the impact of perpendicular flow shear. Specifically, we are primarily interested in the behavior after a few e-folding times, rather than the details of the transients at the very beginning of the evolution. This limit is reached by assuming that $\gamma t \gg 1$ while retaining $|\omega_{V\perp} t| \ll 1$, which implies that $\omega_{V\perp} \ll \gamma$. Thus, the dominant terms are those that are quadratic in time, so the factor of $\left( 1 \pm \gamma t \right)$ becomes $\pm \gamma t$. Moreover, we are interested in unstable modes (i.e. $\gamma$ is real), for which the $\Exp{\left(- i \omega + \gamma \right) t}$ term will dominate. In these limits, we can write $\phi_{0} + \phi_{1} = C_{\gamma} ( 1 + \epsilon_{\alpha} t^{2} ) \Exp{\left(i \omega + \gamma \right) t}$, where
\begin{align}
  \epsilon_{\alpha} = - \frac{1}{4 \gamma t} \left[ \left( - i \omega - \gamma \right)^{2} \frac{\left( k_{\perp}^{2} \rho_{S}^{2} \right)_{1}}{1 + k_{\perp}^{\varnothing 2} \rho_{S}^{2}} - i \left( - i \omega - \gamma \right) \frac{\left( k_{x} \rho_{S} \right)_{1}  \omega_{Mx}}{1 + k_{\perp}^{\varnothing 2} \rho_{S}^{2}} \right]
\end{align}
is a complex constant that can be calculated from equation \refEq{eq:linearSolNextOrder} and the expansion of equation \refEq{eq:hatDef}. To next order in $|\omega_{V\perp} t| \ll 1$, $| \phi_{0} + \phi_{1} |^{2} = |C_{\gamma}|^{2} ( 1 + 2 ~ \text{Real} (\epsilon_{\alpha}) t^{2} ) \Exp{2 \gamma t}$. Thus, the sign of the real part of $\epsilon_{\alpha}$ indicates if perpendicular flow shear will enhance or stabilize the growth of the instability. By manipulating the coefficient, we find that the instability will be enhanced by flow shear if and only if
\begin{align}
   -\frac{k_{||} c_{S} k_{y} \rho_{S} \omega_{V||} - k_{||}^{2} c_{S}^{2}}{1 + k_{\perp}^{\varnothing 2} \rho_{S}^{2}} \frac{\left( k_{\perp}^{2} \rho_{S}^{2} \right)_{1}}{1 + k_{\perp}^{\varnothing 2} \rho_{S}^{2}} + \frac{1}{2} & \left( \frac{K_{x} \rho_{S} \omega_{Mx} + k_{y} \rho_{S} \omega_{My}}{1 + k_{\perp}^{\varnothing 2} \rho_{S}^{2}} \right)^{2} \frac{\left( k_{\perp}^{2} \rho_{S}^{2} \right)_{1}}{1 + k_{\perp}^{\varnothing 2} \rho_{S}^{2}} \label{eq:lowFlowNextOrderCond} \\
   &- \frac{1}{2} \frac{K_{x} \rho_{S} \omega_{Mx} + k_{y} \rho_{S} \omega_{My}}{\left( 1 + k_{\perp}^{\varnothing 2} \rho_{S}^{2} \right)^{2}} \left( k_{x} \rho_{S} \right)_{1} \omega_{Mx} > 0 . \nonumber
\end{align}
We write the condition in this particular form to facilitate comparison with equation \refEq{eq:noFlowGrowthRate}, the instability growth rate. We see that this condition implies that {\it the instability will be enhanced if the finite flow shear correction to the growth rate is positive.} This is an intuitive result and is consistent with the discussion of the $\omega_{V\perp} \ll \gamma$ limit that followed equation \refEq{eq:noFlowGrowthRate}. If flow shear is moving the mode to a wavenumber that is more strongly driven (relative to damping), then it should begin to grow faster. Note that converting from $\hat{\phi}_{1}$ to $\phi_{1}$ using an expanded version of equation \refEq{eq:hatDef} does not end up having an effect on equation \refEq{eq:lowFlowNextOrderCond}. This is because it only introduces terms that are either linear in $t$ or imaginary.

The solution we have derived in this section will only be valid for a short time. Our intuition is that a mode will only be unstable for a finite time before flow shear advects them to high radial wavenumbers, where damping is more effective. However, this hypothesis will be rigorously investigated in the next section.

\subsection{The $|\omega_{V\perp} t| \gg 1$ limit}

We can also obtain analytic solutions to equation \refEq{eq:phiLinearEvolution} in the long time limit. This reveals the ultimate fate of the mode as it is advected to high $k_{x}$ by flow shear, where it can be stabilized by finite sound gyroradius effects. Alternatively, it can be used to understand the behavior of a high $k_{x}$ mode as it is advected towards $k_{x} = 0$. In this limit, equation \refEq{eq:phiLinearEvolution} becomes
\begin{align}
  \left( k_{y}^{2} \rho_{S}^{2} \omega_{V \perp}^{2} t^{2} \right) \left. \frac{\partial^{2} \hat{\phi}}{\partial t^{2}} \right|_{K_{x}} &- i \left( k_{y} \rho_{S} \omega_{Mx} \omega_{V \perp} t \right)  \left. \frac{\partial \hat{\phi}}{\partial t} \right|_{K_{x}} - \left( k_{||} c_{S} k_{y} \rho_{S} \omega_{V||} - k_{||}^{2} c_{S}^{2} \right) \hat{\phi} = 0 . \label{eq:phiLongTimeLinearEvolution}
\end{align}
This equation is solved by a polynomial that, when substituted into equation \refEq{eq:hatDef}, gives
\begin{align}
  \phi = C_{\epsilon} t^{i \tilde{\omega} + \tilde{\gamma} + 1/2 - 2} + C_{\zeta} t^{i \tilde{\omega} - \tilde{\gamma} + 1/2 - 2} ,
\end{align}
where
\begin{align}
  \tilde{\omega} &\equiv - \frac{1}{2} \frac{\omega_{Mx}}{k_{y} \rho_{S} \omega_{V\perp}} \\
  \tilde{\gamma} &\equiv \sqrt{\frac{1}{4} + \frac{k_{||} c_{S} k_{y} \rho_{S} \omega_{V||} - k_{||}^2 c_{S}^{2}}{k_{y}^{2} \rho_{S}^{2} \omega_{V\perp}^2} - \frac{1}{4} \frac{\omega_{Mx}^2}{k_{y}^{2} \rho_{S}^2 \omega_{V\perp}^2} + \frac{i}{2} \frac{\omega_{Mx}}{k_{y} \rho_{S} \omega_{V\perp}}} .
\end{align}
It is important to note that $\tilde{\gamma}$ is a complex number. Additionally, since the time dependence is polynomial rather than exponential as before, the transformation from $\hat{\phi}$ to $\phi$ has been included according to equation \refEq{eq:hatDef}. The finite gyroradius factor has introduced a factor of $t^{-2}$, while the exponential phase factor has changed the sign of $\tilde{\omega}$. From this solution, we can calculate the condition for an unbounded solution as $t \rightarrow \infty$. This condition is simply $\text{Real} (\tilde{\gamma}) > 3/2$, which is equivalent to
\begin{align}
  \omega_{V||} > \frac{1}{k_{||} c_{S} k_{y} \rho_{S}} \left( k_{||}^{2} c_{S}^{2} + 2 k_{y}^{2} \rho_{S}^{2} \omega_{V\perp}^{2} + \frac{2}{9} \omega_{Mx}^{2} \right) \label{eq:unboundedPVGcond}
\end{align}
or
\begin{align}
  \omega_{V||}^{2} > 8 \omega_{V\perp}^{2} + \frac{8}{9} \frac{\omega_{Mx}^{2}}{k_{y}^{2} \rho_{S}^{2}} \label{eq:unboundedPVGcondSimple}
\end{align}
at the most unstable parallel wavenumber of $k_{||} c_{S} = k_{y} \rho_{S} \omega_{V||} / 2$. If $\omega_{V||}$ is greater than this value, then $\phi \rightarrow \infty$ as $t \rightarrow \infty$, indicating that the stabilizing mechanisms are not sufficiently strong to restrain the PVG instability. This condition is verified against the numerical solution to equation \refEq{eq:phiLinearEvolution} in figure \ref{fig:linearResultsAllTime}. Such a condition is interesting as it indicates what mechanisms are able to saturate PVG turbulence. For weak $\omega_{V\perp}$, the nonlinear term can be expected to limit the modes. For strong $\omega_{V\perp}$, the flow shear on its own is capable of damping the modes, without the nonlinear term.

Additionally, equation \refEq{eq:unboundedPVGcondSimple} can be cast into the form of the approximate ``$\omega_{V\perp} \approx \gamma$'' quench rule \cite{WaltzFlowShear1998}, which states that turbulence is quenched when the perpendicular flow shear becomes approximately equal to the maximum linear growth rate $\gamma$ in the absence of perpendicular flow shear. Ignoring the magnetic drifts for simplicity, we can solve equation \refEq{eq:noFlowGrowthRate} for $\omega_{V||}$ using the most unstable parallel wavenumber of $k_{||} c_{S} = k_{y} \rho_{S} \omega_{V||} / 2$. Then we can substitute this $\omega_{V||}$ in order to rewrite equation \refEq{eq:unboundedPVGcondSimple} as
\begin{align}
  \left| \gamma \right| > \frac{\sqrt{2} k_{y} \rho_{S}}{\sqrt{1 + k_{\perp}^{\varnothing 2} \rho_{S}^{2}}} \left| \omega_{V\perp} \right| . \label{eq:turbQuench}
\end{align}
For a typical wavenumber of the most unstable mode, $\vec{K} \rho_{S} = (0, 0.3)$, we find the condition $\left| \omega_{V\perp} \right| > 2.5 \left| \gamma \right|$ for turbulent stabilization. Moreover, the fact that stabilization becomes impossible as $k_{y} \rightarrow 0$ makes sense because perpendicular flow shear only affects a mode in proportion to its $k_{y}$ (as seen in equation \refEq{eq:flowShearTransform}).

\begin{figure}
  \centering
  \includegraphics[width=0.7\textwidth]{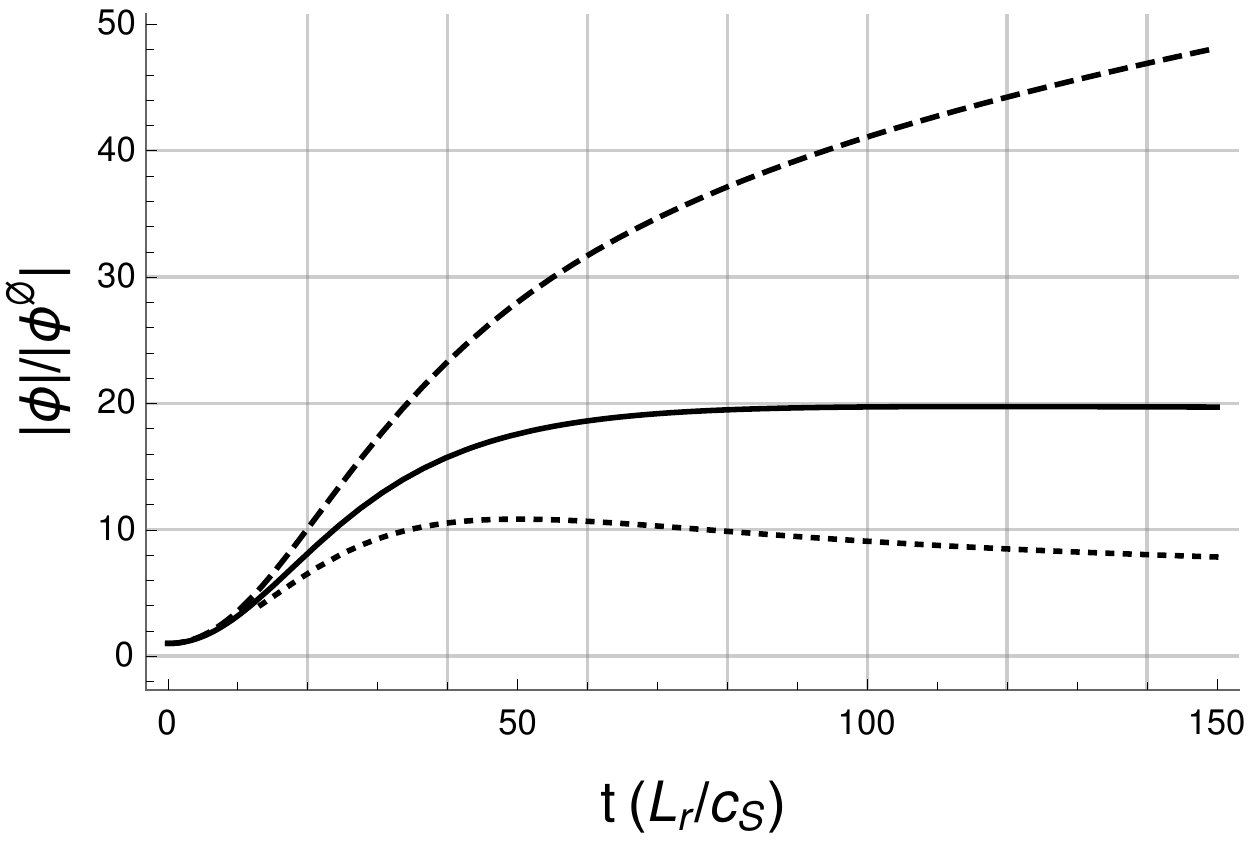}
  \caption{The mode evolution for $\omega_{V||} = \omega_{V||}^{CRIT}$ (solid black), $\omega_{V||} = 1.05 \omega_{V||}^{CRIT}$ (dashed black), and $\omega_{V||} = 0.95 \omega_{V||}^{CRIT}$ (dotted black), where $\omega_{V||}^{CRIT} = 1.67 c_{S}/L_{r}$ is determined by equation \refEq{eq:unboundedPVGcondSimple}. All other parameters are identical to figure \ref{fig:linearResultsShortTime}.}
  \label{fig:linearResultsAllTime}
\end{figure}

Notice that equation \refEq{eq:unboundedPVGcond} is fairly similar to equation \refEq{eq:PVGstabilityShortTime}, the condition for instability at $t=0$. The primary difference is that the stabilizing perpendicular flow shear term appears and the stability caused by the $\omega_{Mx}$ magnetic drift term slightly weakens. Interestingly, this indicates that for large $K_{x}$, small (but non-zero) $\omega_{V\perp}$, and/or large $\omega_{My}/\omega_{Mx}$, the mode can be initially stable, but unbounded at long time after being advected by flow shear. Moreover, due to the finite sound gyroradius effect contained in equation \refEq{eq:hatDef}, the mode can actually decay quadratically, but then have an unbounded limit. In fact, as shown in figure \ref{fig:stableButUnbounded}, the time dependence can be even more complex. Initially, the mode is stable according to equation \refEq{eq:noFlowGrowthRate} and its amplitude decreases due to finite sound gyroradius damping as its radial wavenumber changes (i.e. equation \refEq{eq:hatDef}). Shortly thereafter, other effects of flow shear become large enough to drive the mode unstable according to equation \refEq{eq:linearSolNextOrder}. Then, at intermediate times, both components of the magnetic drifts are important, which reduces the mode growth sufficiently for the finite sound gyroradius damping to win again. Lastly, at long times, the $\omega_{My}$ term becomes negligible, allowing the parallel flow shear to slightly overpower $\omega_{Mx}$ and the finite sound gyroradius damping according to equation \refEq{eq:unboundedPVGcondSimple}. Thus, the mode grows polynomially without limit.

\subsection{Zonal modes}

Lastly, for zonal modes (i.e. $k_{y}=0$), we can directly solve equation \refEq{eq:phiLinearEvolution} and find a simple solution without needing to take any limits. This is because the effects of both parallel and perpendicular flow shear are proportional to $k_{y}$, so they vanish. We are left with a simple 2\textsuperscript{nd} order ordinary differential equation with constant coefficients, just like in section \ref{subsec:PVGlowFlowLimit}. Thus, the solution can be found by setting $k_{y} = 0$ in equation \refEq{eq:noFlowSol} to get
\begin{align}
  \phi_{Z} =& C_{\eta} \Exp{-i \left(\frac{1}{2} \frac{K_{x} \rho_{S} \omega_{Mx}}{1 + K_{x}^{2} \rho_{S}^{2}} + \sqrt{\frac{k_{||}^{2} c_{S}^{2}}{1 + K_{x}^{2} \rho_{S}^{2}} + \frac{1}{4} \left( \frac{K_{x} \rho_{S} \omega_{Mx}}{1 + K_{x}^{2} \rho_{S}^{2}} \right)^{2}} \right) t} \\
  &+ C_{\theta} \Exp{-i \left(\frac{1}{2} \frac{K_{x} \rho_{S} \omega_{Mx}}{1 + K_{x}^{2} \rho_{S}^{2}} - \sqrt{\frac{k_{||}^{2} c_{S}^{2}}{1 + K_{x}^{2} \rho_{S}^{2}} + \frac{1}{4} \left( \frac{K_{x} \rho_{S} \omega_{Mx}}{1 + K_{x}^{2} \rho_{S}^{2}} \right)^{2}} \right) t} . \nonumber
\end{align}
We see that the zonal modes are not driven nor damped linearly \cite{RosenbluthZonalFlowDamping1998} and oscillate if they have a finite $k_{||}$ and/or there is a magnetic drift in the $x$ direction. Like other gradient-driven instabilities, the PVG cannot drive the zonal modes directly because its effect is proportional to $k_{y}$. Instead the zonal modes must be driven by the nonlinear interaction of pairs of non-zonal modes (the focus of section \ref{subsec:twoPumpModes}).

\section{Nonlinear two-dimensional slab results}
\label{sec:nonlinearResults}

To investigate the nonlinear dynamics of turbulence, we will let $k_{||} = 0$ in order to study two-dimensional turbulence. Strictly speaking, this is inconsistent with our assumption of adiabatic electrons \cite{DominskiElectronResponse2015}. However, the dynamics should be similar to a very small value of $k_{||}$ and setting $k_{||}$ equal to exactly zero facilitates benchmarking against gyrokinetic codes. In this section, we will focus on how perpendicular flow shear alters the fundamental nonlinear drive of small amplitude ``driven'' modes by large ``pump'' modes. The analysis will be kept general for benchmarking purposes, but the primary physics application is zonal flow dynamics. Zonal flows cannot be directly driven by background gradients, so they only grow due to the nonlinear coupling of non-zonal modes. Nevertheless, in many if not most gyrokinetic simulations, the amplitude of the zonal flows eventually become much larger than the non-zonal modes \cite{DimitsShift2000, RogersZonalFlows2000}. Thus, despite the fact that zonal modes do not directly cause transport, the mechanism by which they regulate non-zonal modes is of much interest.

With $k_{||} = 0$, the evolution equation for $\phi$ (i.e. equation \refEq{eq:phiEvolutionFinal}) can be formulated as
\begin{align}
  \left. \frac{\partial}{\partial t} \right|_{K_{x}} \left( \left( 1 + k^{2} \rho_{S}^{2} \right) \phi \right) &+ \frac{1}{2 B} \sum_{\vec{K}'} \left( \vec{K}' \times \vec{K}'' \right) \cdot \hat{b} \left( k'^{2} - k''^{2} \right) \rho_{S}^{2} \phi'^{\ast} \phi''^{\ast} \label{eq:phi2DEvolution} \\
  &+ i \left( k_{x} \rho_{S} \omega_{Mx} + k_{y} \rho_{S} \omega_{My} \right) \phi = 0 , \nonumber
\end{align}
where the coupling condition is $\vec{K} + \vec{K}' + \vec{K}'' = 0$. Note that we have redefined $\vec{K}' \rightarrow - \vec{K}'$ and $\vec{K}'' \rightarrow - \vec{K}''$ and used the reality condition that $\phi ( - \vec{K} ) = \phi^{\ast} ( \vec{K} )$ (where the $\ast$ superscript indicates the complex conjugate). We see that the dependence on the parallel velocity moment and the parallel velocity gradient has dropped out. This is not too troubling because the parallel flow shear does not enter into the nonlinear term, which is the primary focus of our analysis. We note that equation \refEq{eq:phi2DEvolution} is a generalization of the Charney-Hasegawa-Mima equation \cite{HasegawaMimaEquation1978} to include the perpendicular flow shear and simple magnetic drifts.

Substituting the hat notation from equation \refEq{eq:hatDef} yields
\begin{align}
  \left. \frac{\partial \hat{\phi}}{\partial t} \right|_{K_{x}} = \frac{1}{2} \sum_{\vec{K}'} & \Lambda(\vec{K}, \vec{K}', \vec{K}'', t) ~ \hat{\phi}'^{\ast} \hat{\phi}''^{\ast} \label{eq:phiEvolutionHat} \\
  &\times \frac{1 + k^{2} \rho_{S}^{2}}{(1 + k'^{2} \rho_{S}^{2}) (1 + k''^{2} \rho_{S}^{2})} \Exp{i \left. \int \right|_{K_{x}} dt ~ \theta (\vec{K}, \vec{K}', \vec{K}'', t)} , \nonumber
\end{align}
where we have defined
\begin{align}
  \Lambda(\vec{K}, \vec{K}', \vec{K}'', t) &\equiv - \frac{1}{B} \left( \vec{K}' \times \vec{K}'' \right) \cdot \hat{b} ~ \frac{\left( k'^{2} - k''^{2} \right) \rho_{S}^{2}}{1 + k^{2} \rho_{S}^{2}} \label{eq:lambdaDef} \\
  W ( \vec{K} , t ) &\equiv \frac{k_{x} \rho_{S} \omega_{Mx} + k_{y} \rho_{S} \omega_{My}}{1 + k^{2} \rho_{S}^{2}} \label{eq:Wdef} \\
  \theta(\vec{K}, \vec{K}', \vec{K}'', t) &\equiv W ( \vec{K}, t ) + W ( \vec{K}', t ) + W ( \vec{K}'', t ) . \label{eq:thetaDef}
\end{align}

Now, as in a three-wave resonant decay calculation \cite{HasegawaThreeWaveCoupling1979}, we will consider three Fourier modes with $\vec{K}_{a} + \vec{K}_{b} + \vec{K}_{c} = 0$ such that they nonlinearly couple. We will consider two separate cases: two pump modes such that $|\phi_{b}| \sim |\phi_{c}| \gg |\phi_{a}|$ and a single pump mode such that $|\phi_{b}| \gg |\phi_{c}| \sim |\phi_{a}|$.  Here and henceforth the Latin letter subscript(s) specifies the Fourier modes of the wavenumber arguments. The first case is straightforward to solve analytically and can model the initial drive of the zonal modes from zero amplitude. The second case is more complex, but is important for understanding how dominant zonal modes couple to non-zonal modes.

\subsection{Two pump mode case}
\label{subsec:twoPumpModes}

Linearizing equation (\ref{eq:phiEvolutionHat}) in the ratio of the amplitude of the driven mode ``a'' to the amplitudes of the two pump modes ``b'' and ``c'' produces
\begin{align}
  \left. \frac{\partial \hat{\phi}_{a}}{\partial t} \right|_{K_{x}} &= \Lambda_{abc} ~ \hat{\phi}^{\ast}_{b} \hat{\phi}^{\ast}_{c} \frac{1 + k_{a}^{2} \rho_{S}^{2}}{(1 + k_{b}^{2} \rho_{S}^{2}) (1 + k_{c}^{2} \rho_{S}^{2})} \Exp{i \left. \int \right|_{K_{x}} dt ~ \theta} \label{eq:mode1evolutionTwoPump} \\
  \left. \frac{\partial \hat{\phi}_{b}}{\partial t} \right|_{K_{x}} &= 0 \label{eq:mode2evolutionTwoPump} \\
  \left. \frac{\partial \hat{\phi}_{c}}{\partial t} \right|_{K_{x}} &= 0 , \label{eq:mode3evolutionTwoPump}
\end{align}
where the mode subscripts have been dropped from $\theta$ because their order has no effect. Immediately, we see that the pump modes have the solutions of $\hat{\phi}_{b} = \hat{\phi}_{b}^{\varnothing}$ and $\hat{\phi}_{c} = \hat{\phi}_{c}^{\varnothing}$. This result is consistent with our previous analysis, as it satisfies equation \refEq{eq:phiLinearEvolution} when $k_{||} = 0$. With a constant $\hat{\phi}_{b}$, equation \refEq{eq:hatDef} tells us that
\begin{align}
  \phi_{b} = \phi_{b}^{\varnothing} \frac{1 + k_{b}^{\varnothing 2} \rho_{S}^{2}}{1 + k_{b}^{2} \rho_{S}^{2}} \Exp{- i \left. \int \right|_{K_{x}} dt ~ W_{b}} \label{eq:pumpModeSol}
\end{align}
and the same is true for mode ``c''. Hence, we see that the phase of each pump mode oscillates with a non-constant frequency due to the gradients in the plasma. The only effect that alters the magnitude of the modes is the perpendicular flow shear, which enters through the factor $( 1 + k_{\perp}^{\varnothing 2} \rho_{S}^{2} ) / ( 1 + k_{\perp}^{2} \rho_{S}^{2} )$. This advects the mode in $k_{x}$ and modifies the finite gyroradius effects that originated in the quasineutrality equation. However, if a pump mode is zonal the wavenumber is not sheared, so that the mode maintains a constant amplitude. Lastly, we note that, because we have linearized our differential equations, our analytic calculations will not exhibit pump depletion (i.e. losses due to the energy transferred to the driven mode). However, pump depletion is negligible as long as the amplitude of the driven mode remains much smaller than the pump modes.

\begin{figure}
  \centering
  \includegraphics[width=0.65\textwidth]{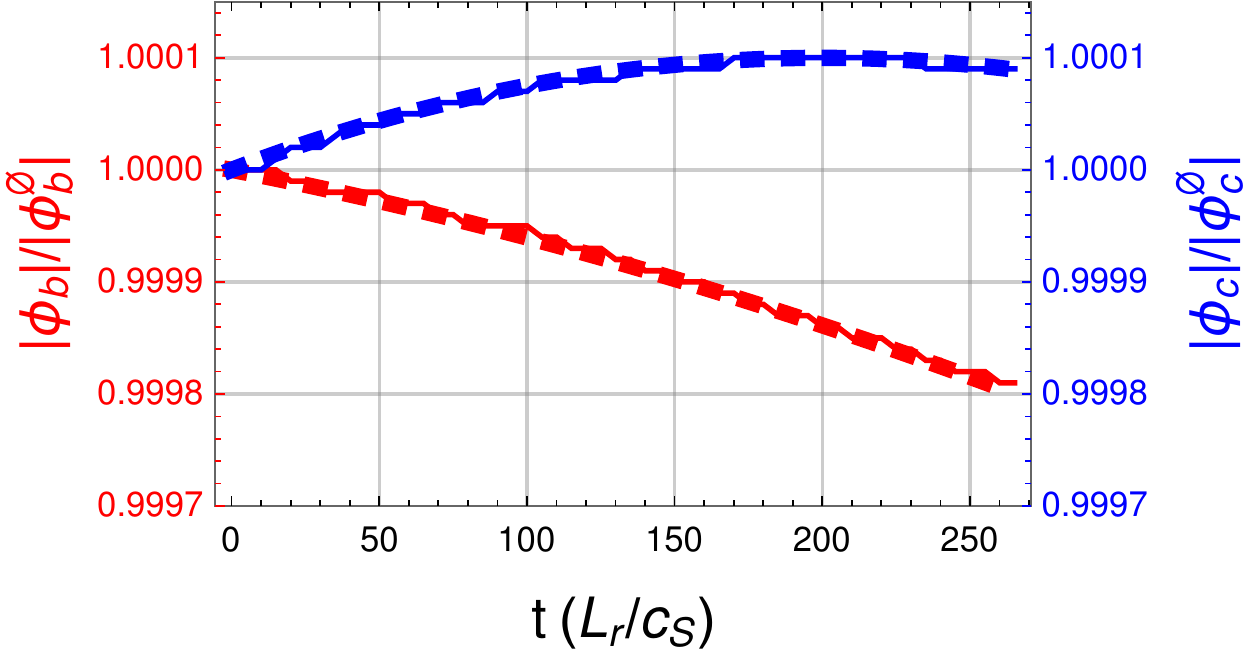}
  \includegraphics[width=0.56\textwidth]{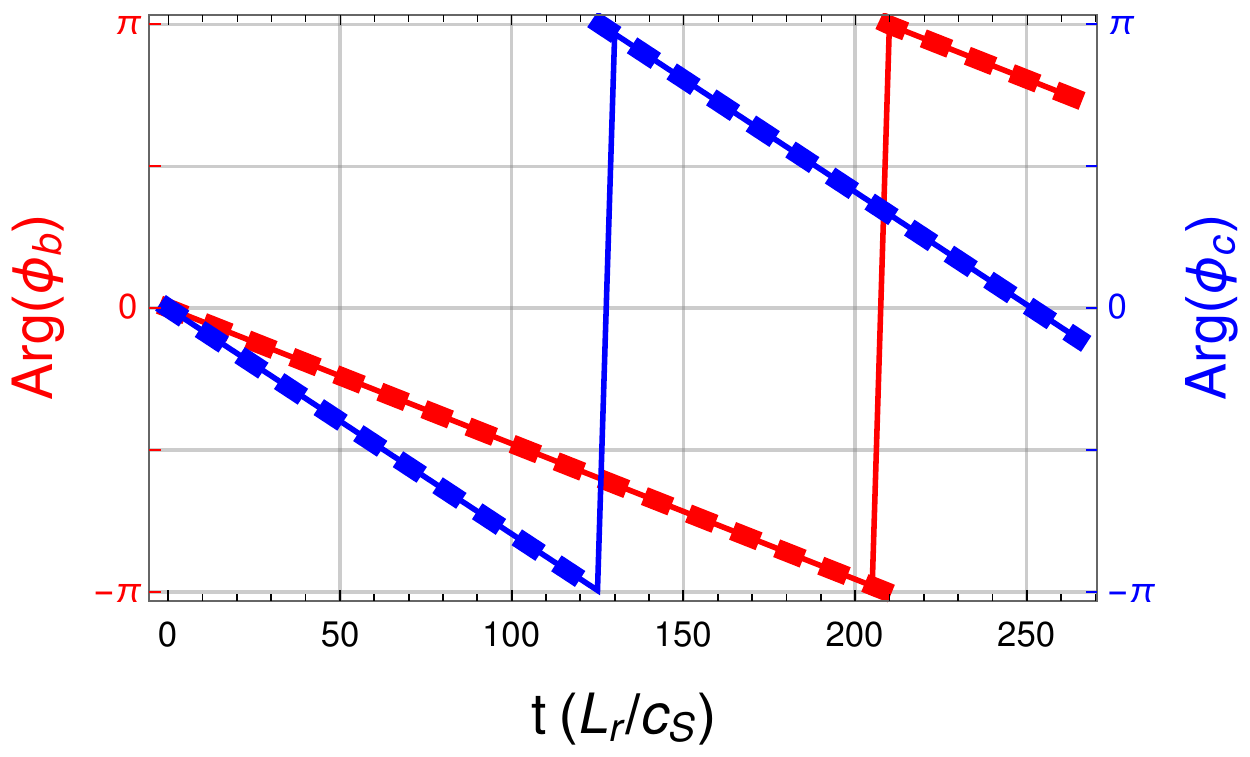}
  \caption{The amplitude (top) and phase (bottom) of the two pump modes ``b'' (red) and ``c'' (blue) calculated by equation (\ref{eq:pumpModeSol}) (thick dashed) and GENE (thin solid) for the parameters given in the text.}
  \label{fig:twoPumpPump}
\end{figure}

Since $\hat{\phi}_{b}$ and $\hat{\phi}_{c}$ are known, equation \refEq{eq:mode1evolutionTwoPump} can be integrated directly to find the evolution of the driven mode. Substituting the form of $\Lambda_{abc}$ gives
\begin{align}
   \hat{\phi}_{a} &= - \hat{\phi}^{\ast}_{b} \hat{\phi}^{\ast}_{c} \frac{1}{B} \left( \vec{K}_{b} \times \vec{K}_{c} \right) \cdot \hat{b} \left. \int \right|_{K_{x}} dt \left[ \frac{\left( k_{b}^{2} - k_{c}^{2} \right) \rho_{S}^{2}}{(1 + k_{b}^{2} \rho_{S}^{2}) (1 + k_{c}^{2} \rho_{S}^{2})} \Exp{i \left. \int \right|_{K_{x}} dt ~ \theta} \right] . \label{eq:mode1SolTwoPump}
\end{align}
Thus, we see that for pump modes with fixed amplitudes $\hat{\phi}_{b}$ and $\hat{\phi}_{c}$, the driven mode ``a'' does not grow exponentially with time, it can only grow polynomially. This is consistent with the numerical results of reference \cite{RogersZonalFlows2000}, which used a realistic toroidal geometry. Equation \refEq{eq:mode1SolTwoPump} also shows that, if the three modes are oscillating out of phase (i.e. $\theta \neq 0$), then the solution will be oscillatory. However, the presence of perpendicular flow shear (i.e. $\omega_{V\perp} \neq 0$) causes $\theta$ to vary with time. Thus, $\theta ( t )$ can pass through $\theta = 0$, allowing the mode to grow temporarily. When $\omega_{V\perp} \neq 0$ and $\theta \approx 0$, $k_{x}$ varies linearly in time for all non-zonal modes, which means that temporarily $\left| \phi_{a} \right|$ can grow quickly (e.g. if $k^{2} \rho_{S}^{2} \ll 1$, $\left| \phi_{a} \right|$ can grow as $t^{3}$). The condition of $\theta \approx 0$ is equivalent to the frequency matching condition that is required for mode growth in three-mode coupling without background flow shear \cite{HasegawaThreeWaveCoupling1979}. For example, when $\omega_{V\perp} = 0$, the wavenumbers in equation \refEq{eq:mode1SolTwoPump} become constant and $|\phi_{a}|$ can grow linearly with $t$, but only if $\theta = 0$.

\begin{figure}
	\centering
	\includegraphics[width=0.65\textwidth]{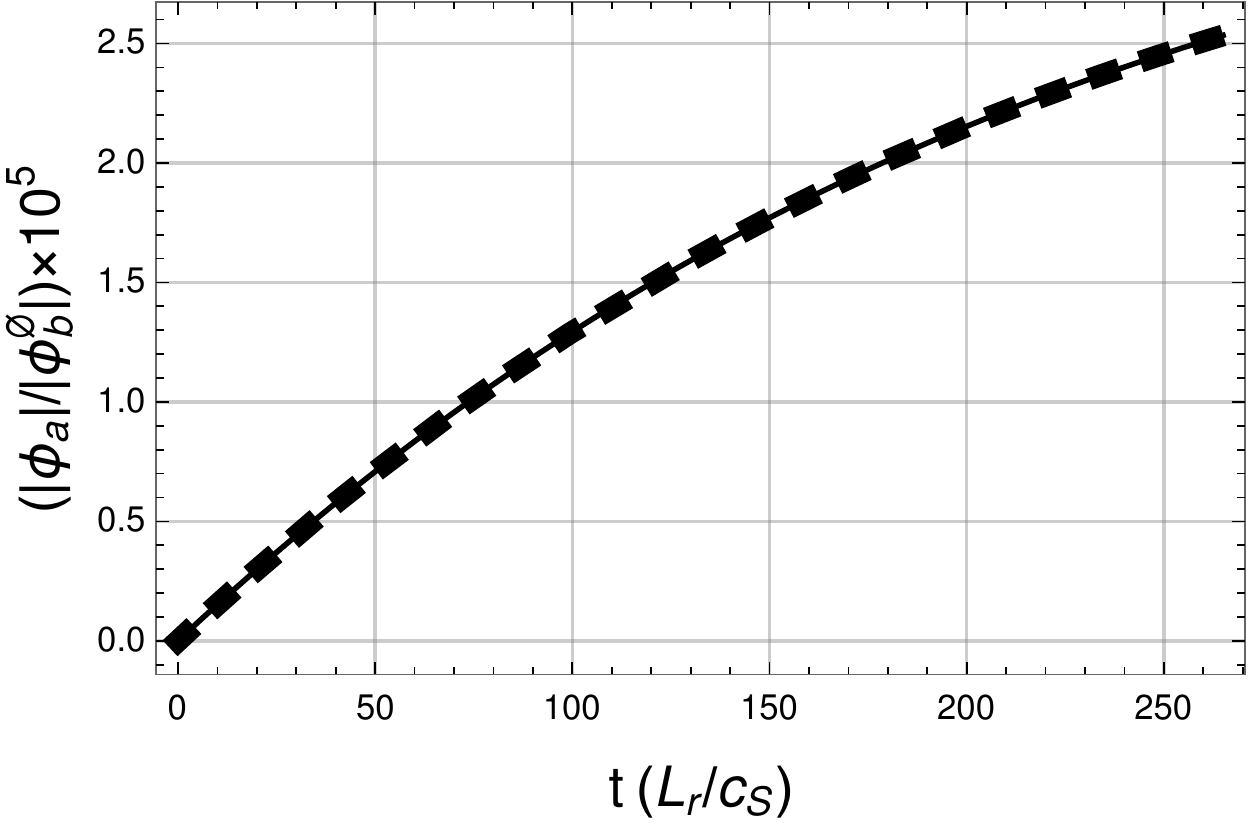}
	\includegraphics[width=0.65\textwidth]{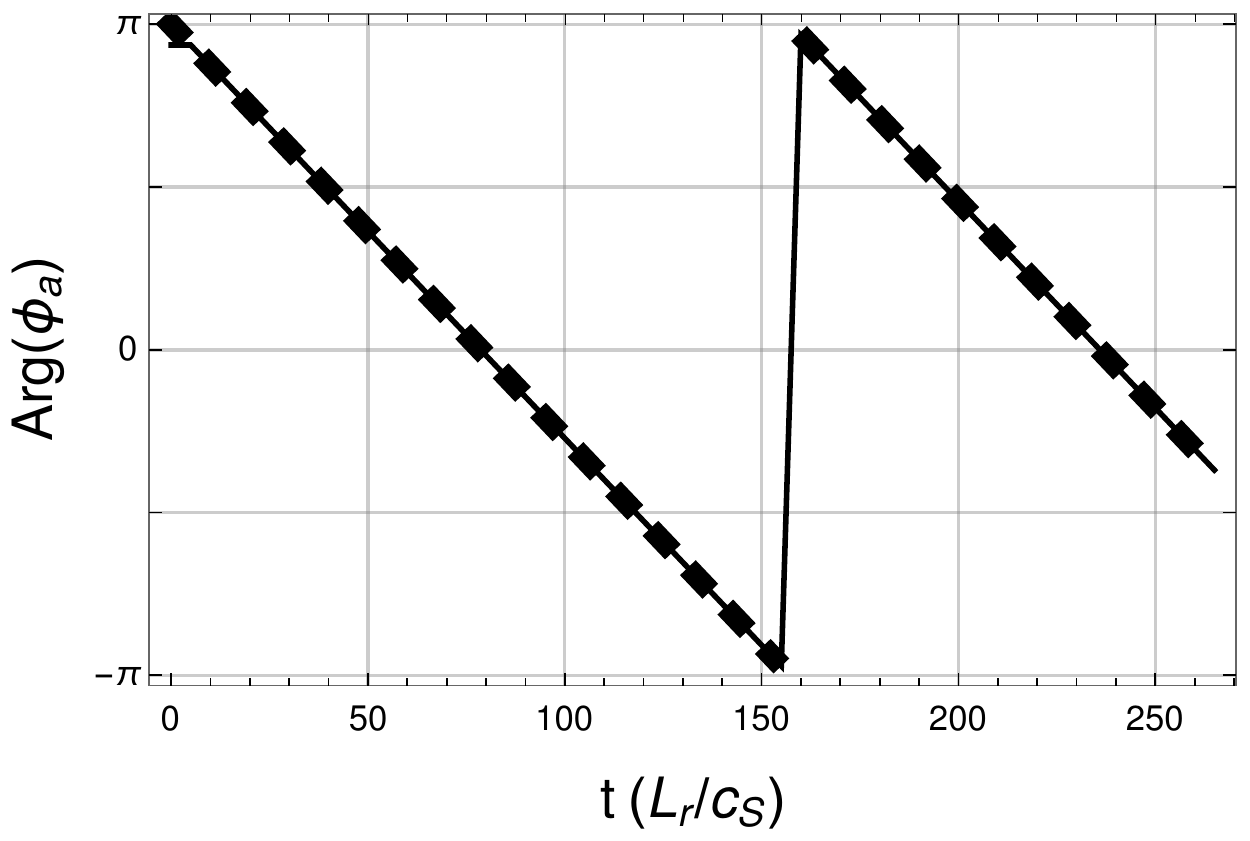}
	\caption{The amplitude (top) and phase (bottom) of the driven mode calculated by equations \refEq{eq:hatDef} and \refEq{eq:mode1SolTwoPump} (thick dashed) and GENE (thin solid) for the parameters given in the text.}
	\label{fig:twoPumpDriven}
\end{figure}

In the long time limit, the nonlinear drive (i.e. the right side of equation \refEq{eq:mode1SolTwoPump}) for a non-zonal mode ``a'' is proportional to $t^{-1}$. This means that $\left| \phi_{a} \right|$ decays away to zero as $t^{-3}$. However, the situation of most interest is when mode ``a'' is zonal. When $k_{ay} = 0$, the coupling condition requires the other two modes to have the same value of $k_{y}$, so they get sheared at the same rate. Thus, the numerator of equation \refEq{eq:mode1SolTwoPump} varies linearly in time instead of quadratically and the drive of a zonal mode decays {\it more} quickly than a non-zonal one (specifically as $t^{-2}$ instead of $t^{-1}$). Lastly, when one of the pump modes is zonal, the drive actually gets stronger with time (specifically it scales linearly with $t$).

These analytic solutions are again verified by comparison with the gyrokinetic code GENE as shown in figures \ref{fig:twoPumpPump} and \ref{fig:twoPumpDriven}. The simulation is initialized with $\hat{\phi}_{b}^{\varnothing}/ (B \rho_{S}^{2}) = \hat{\phi}_{c}^{\varnothing}/ (B \rho_{S}^{2}) = c_{S}/L_{r}$, while all other modes are zero. The wavenumbers are chosen to be $\vec{K}_{b} \rho_{S} = (-0.0086,0.015)$ and $\vec{K}_{c} \rho_{S} = (0.01,0.025)$, while $\omega_{V\perp} = -0.002 c_{S}/L_{r}$, $\omega_{Mx} = 0$, and $\omega_{My} = c_{S}/L_{r}$. Figure \ref{fig:twoPumpPump} shows that the pump mode ``b'' decreases because flow shear advects it to large $k_{\perp}^{2}$, where the finite sound gyroradius effects more effectively average over it. The pump mode ``c'' increases briefly because flow shear initially advects it to a lower value of $k_{\perp}^{2}$, but then it starts to decay after passing through $k_{x} = 0$. The only other mode that grows is the one with $\vec{K}_{a} = - (\vec{K}_{b} + \vec{K}_{c})$, which is driven by the nonlinear coupling and is shown in figure \ref{fig:twoPumpDriven}.

Finally, note that the calculation in this section for the initial drive of {\it zonal} modes can be extended to more than just three modes. This is because, during the initial linear phase of the drive, the evolution of all the non-zonal modes is determined by the linear physics (i.e. the drive from the background gradients). Moreover, the zonal modes do not directly couple with each other nonlinearly because their wavevectors are parallel to each other such that $\vec{K}' \times \vec{K}'' = 0$ in equation \refEq{eq:lambdaDef}. This means that the non-zonal modes can be solved without any nonlinearity and the evolution of each zonal mode can be calculated from equation \refEq{eq:mode1SolTwoPump} with a summation over all non-zonal modes out in front. Because the drive of zonal modes decays more quickly with time, their drive will tend to be more dominated by the modes near $k_{x} = 0$ (compared to the nonlinear drive of non-zonal modes).

\subsection{One pump mode case}

In this section, instead of having two large amplitude pump modes, we will include just one: $|\phi_{b}| \gg |\phi_{a}| \sim |\phi_{c}|$. This is appropriate for modeling parametric instability decay. While the previous section studied the initial growth of a zonal mode from small amplitude, this section will focus on the opposite situation. In the quasi-steady-state of nonlinearly saturated turbulence, the amplitudes of the zonal modes are typically much larger than that of the non-zonal modes \cite{DimitsShift2000, RogersZonalFlows2000}. Thus, we would like to study how a single large-amplitude zonal mode regulates small-amplitude non-zonal modes. However, we will keep the analysis general to allow for a non-zonal pump mode.

\begin{figure}
  \centering
  \includegraphics[width=0.65\textwidth]{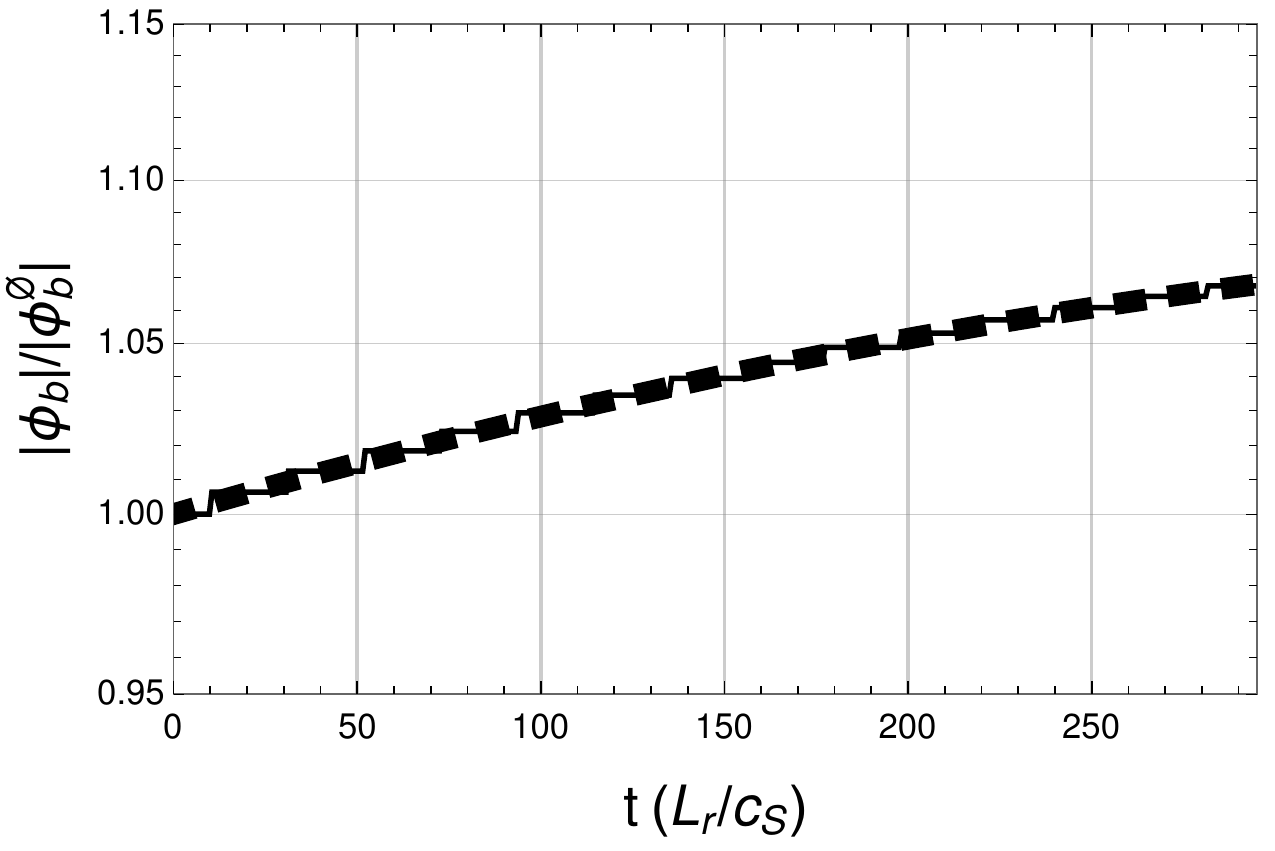}
  \caption{The amplitude of the pump mode as calculated by equation (\ref{eq:pumpModeSol}) (thick dashed) and GENE (thin solid). The parameters used were $\phi_{b}^{\varnothing}/ (B \rho_{S}^{2}) = 8.5 c_{S}/L_{r}$, $\vec{K}_{b} \rho_{S} = (0.3,0.3)$, $\omega_{V\perp} = -0.002 c_{S}/L_{r}$, and $\omega_{Mx} = \omega_{My} = \theta = 0$.}
  \label{fig:onePumpPumpShear}
\end{figure}

Linearizing equation (\ref{eq:phiEvolutionHat}) in the ratio of the driven mode amplitudes (here modes ``a'' and ``c'') to the pump mode amplitude (here mode ``b'') gives
\begin{align}
  \left. \frac{\partial \hat{\phi}_{a}}{\partial t} \right|_{K_{x}} &= \Lambda_{abc} ~ \hat{\phi}^{\ast}_{b} \hat{\phi}^{\ast}_{c} \frac{1 + k_{a}^{2} \rho_{S}^{2}}{(1 + k_{b}^{2} \rho_{S}^{2}) (1 + k_{c}^{2} \rho_{S}^{2})} \Exp{i \left. \int \right|_{K_{x}} dt ~ \theta} \label{eq:mode1evolution} \\
  \left. \frac{\partial \hat{\phi}_{b}}{\partial t} \right|_{K_{x}} &= 0 \label{eq:mode2evolution} \\
  \left. \frac{\partial \hat{\phi}_{c}}{\partial t} \right|_{K_{x}} &= \Lambda_{cab} ~ \hat{\phi}^{\ast}_{a} \hat{\phi}^{\ast}_{b} \frac{1 + k_{c}^{2} \rho_{S}^{2}}{(1 + k_{a}^{2} \rho_{S}^{2}) (1 + k_{b}^{2} \rho_{S}^{2})} \Exp{i \left. \int \right|_{K_{x}} dt ~ \theta} . \label{eq:mode3evolution}
\end{align}
The pump mode solution is again given by equation \refEq{eq:hatDef} with $\hat{\phi}_{b} = \hat{\phi}_{b}^{\varnothing}$, which is compared to GENE simulations in figure \ref{fig:onePumpPumpShear}. We see good agreement, except for the subtle step-like behavior in the GENE evolution (which can also be seen in figure \ref{fig:twoPumpPump}). This is an artifact of the ``wavevector-remap'' scheme used to implement flow shear and will converge to a continuous evolution in the limit of high radial wavenumber resolution \cite{McmillanRemap2017}.

Note that strictly speaking equations \refEq{eq:mode1evolution} and \refEq{eq:mode3evolution} should have a second term. This is because, due to the reality condition of $\phi ( - \vec{K} ) = \phi^{\ast} ( \vec{K} )$, the pump mode is actually two modes --- one at $\vec{K}_{b}$ and one at $- \vec{K}_{b}$. Thus, there is a second set of three modes involving a new mode $\vec{K}_{d}$ that satisfies the coupling condition $\vec{K}_{a} - \vec{K}_{b} + \vec{K}_{d} = 0$. However, this second term can be ignored if it does not lead to instability or if mode ``d'' is not included in the simulation domain.

\begin{figure}
  \begin{adjustbox}{addcode={
  	\begin{minipage}{\width}}{
  	  \caption{The time evolution of driven modes with $K_{ax} \rho_{S} \in [-0.35, -0.1]$ and $k_{ay} \rho_{S} \in [0, 0.3]$ (indicated in the upper left corner of each plot) as calculated by a numerical solution to equation (\ref{eq:drivenModeEvolution}) (thick dashed), the short-time analytic solution of equation \refEq{eq:solNoFlow} (thin dotted), and GENE (thin solid). The parameters used are identical to figure \ref{fig:onePumpPumpShear}. Gray plots indicate that the numerical solution is not expected to match GENE as explained in the text.}
  	  \label{fig:onePumpDrivenShear}
    \end{minipage}},rotate=90,center}
    \includegraphics[scale=0.95]{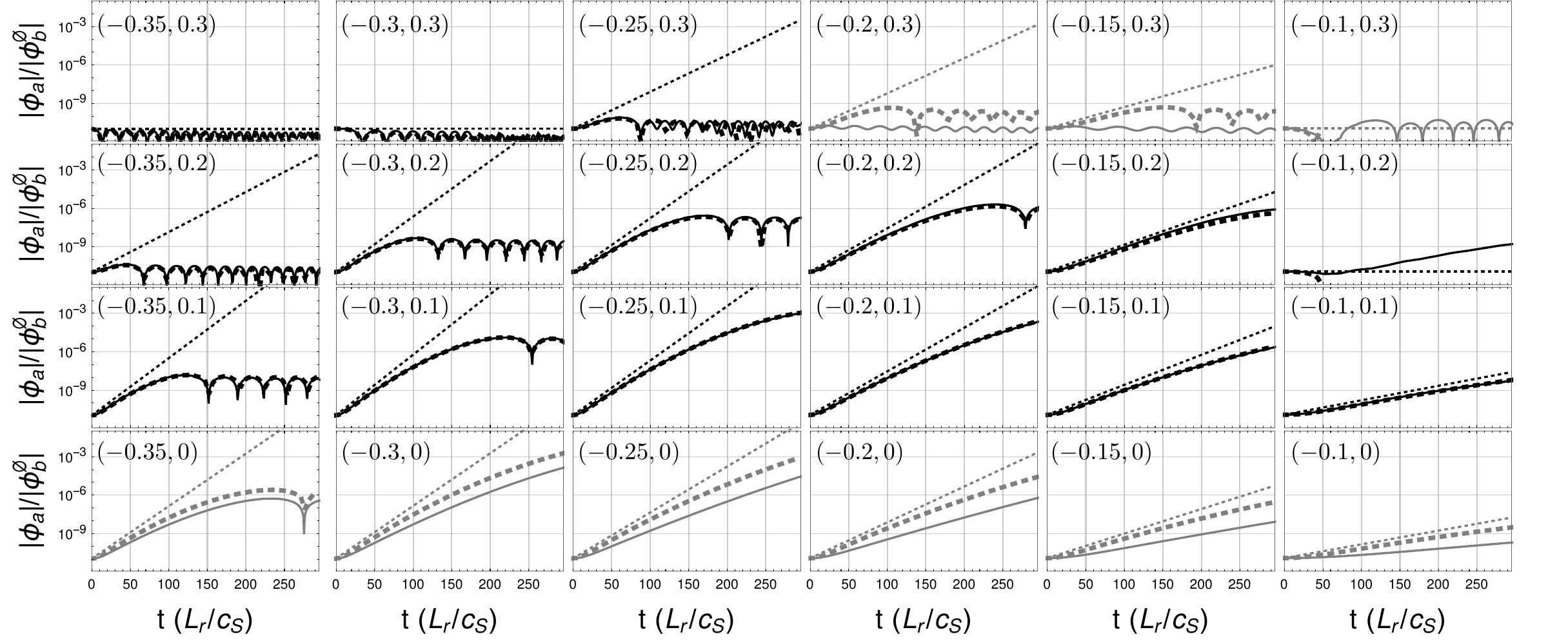}
  \end{adjustbox}
\end{figure}

Calculating the evolution of the driven modes ``a'' and ``c'' requires considerable mathematics. Solving equation \refEq{eq:mode1evolution} for $\hat{\phi}^{\ast}_{c}$ and substituting it into the complex conjugate of equation \refEq{eq:mode3evolution} yields
\begin{align}
  \left. \frac{\partial^{2} \hat{\phi}_{a}}{\partial t^{2}} \right|_{K_{x}} + \left( -i \theta + G_{abc} \right) \left. \frac{\partial \hat{\phi}_{a}}{\partial t} \right|_{K_{x}} - \Lambda_{abc} \Lambda_{cab} \left| \phi_{b} \right|^{2} \hat{\phi}_{a} = 0 , \label{eq:drivenModeEvolution}
\end{align}
where
\begin{align}
  G_{abc} &\equiv \left. \frac{\partial}{\partial t} \right|_{K_{x}} \Ln{\frac{1}{\Lambda_{abc}} \frac{\left( 1 + k_{b}^{2} \rho_{S}^{2} \right) \left( 1 + k_{c}^{2} \rho_{S}^{2} \right)}{1 + k_{a}^{2} \rho_{S}^{2}}} \label{eq:Gdef} \\
  &= 2 \omega_{V\perp} \left( - \frac{k_{bx} k_{by} - k_{cx} k_{cy}}{k_{b}^{2} - k_{c}^{2}} + \frac{k_{bx} k_{by} \rho_{S}^{2}}{1 + k_{b}^{2} \rho_{S}^{2}} + \frac{k_{cx} k_{cy} \rho_{S}^{2}}{1 + k_{c}^{2} \rho_{S}^{2}} \right) \nonumber
\end{align}
and we note that $\left| \phi_{b} \right|^{2}$ can be time dependent. The analogous equation for $\hat{\phi}_{c}$ can be found by swapping the subscripts ``a'' and ``c''. These equations are identical to those in reference \cite{HasegawaThreeWaveCoupling1979}, except for the presence of the flow shear $\omega_{V\perp}$ introduces $G_{abc}$ and makes $\theta$, $\Lambda_{abc}$, and $\left| \phi_{b} \right|^{2}$ depend on time.

Equation \refEq{eq:drivenModeEvolution} is straightforward to solve numerically as it can be written as a 2\textsuperscript{nd} order, ordinary differential equation with polynomial coefficients. This is done for many different ``a'' modes in figure \ref{fig:onePumpDrivenShear}, which are compared against GENE simulations. We see that some modes are unstable for the entire simulation (e.g. $\vec{K}_{a} \rho_{S} = (-0.15,0.1)$), while others only ever oscillate (e.g. $\vec{K}_{a} \rho_{S} = (-0.35,0.3)$). Certain modes start off unstable, but then the perpendicular flow shear advects them to wavenumbers that are stable (e.g. $\vec{K}_{a} \rho_{S} = (-0.25,0.2)$). However, GENE and our numerical solution do not agree well for several modes. This is because the evolution of these modes is not governed by simple three-wave resonant decay. For these cases, there is a fourth mode in the simulation domain that is unstable, which complicates our analysis as discussed at the beginning of this subsection. These cases have been indicated in figure \ref{fig:onePumpDrivenShear} with gray and should not necessarily agree well. All the other modes match the numerical prediction. The most significant source of error was the ``wavevector-remap'' scheme used to model perpendicular flow shear, which is only exact in the limit of infinite $k_{x}$ resolution \cite{McmillanRemap2017}. This error was minimized by using a $k_{x}$ resolution four times denser than the points shown in figure \ref{fig:onePumpDrivenShear}. Note the $\vec{K}_{a} \rho_{S} = (-0.1,0.2)$ mode shows good agreement, but the numerical solution is only plotted for a short time because the numerical solution encountered convergence issues.

Equation \refEq{eq:drivenModeEvolution} does {\it not} have an analytic solution in general because the polynomials are of very high degree (i.e. the effective coefficient for the second derivative term is degree 10, the first derivative term is degree 9, and the $\hat{\phi}_{a}$ term is degree 6). Nevertheless, we see that if the nonlinear coupling is weak, then $\hat{\phi}_{a} = \text{const}$ is a solution. This is an important and nontrivial solution because the mode will still evolve through the dependences in equation \refEq{eq:hatDef}. As in the linear analysis of section \ref{sec:linearResults}, further analytic results can be found by investigating equation \refEq{eq:drivenModeEvolution} in the $|\omega_{V\perp} t| \ll 1$ and $|\omega_{V\perp} t| \gg 1$ limits.

\subsubsection{The $|\omega_{V\perp} t| \ll 1$ limit}
\label{sec:shortTimeNonlinear}

To lowest order in the short time limit, equation \refEq{eq:drivenModeEvolution} becomes
\begin{align}
  \left. \frac{\partial^{2} \hat{\phi}_{a0}}{\partial t^{2}} \right|_{K_{x}} - i \theta_{0} \left. \frac{\partial \hat{\phi}_{a0}}{\partial t} \right|_{K_{x}} - \left( \Lambda_{abc} \Lambda_{cab} \left| \phi_{b} \right|^{2} \right)_{0} \hat{\phi}_{a0} = 0 , \label{eq:drivenModeEvolutionLowestOrder}
\end{align}
where the numerical subscript indicates the quantity's order in the $|\omega_{V\perp} t| \ll 1$ expansion. To lowest order, all the coefficients are constant in time and simply equal to their values without perpendicular flow shear. Surprisingly, this differential equation has the same form as the one we solved in section \ref{subsec:PVGlowFlowLimit}. Using that solution, we find
\begin{align}
  \hat{\phi}_{a0} = C_{\iota} \Exp{\left( i \omega + \gamma \right) t} + C_{\kappa} \Exp{\left( i \omega - \gamma \right) t} , \label{eq:solNoFlow}
\end{align}
where
\begin{align}
  \omega &\equiv \frac{\theta_{0}}{2} \\
  \gamma &\equiv \sqrt{\left( \Lambda_{abc} \Lambda_{cab} \left| \phi_{b} \right|^{2} \right)_{0} - \frac{\theta_{0}^{2}}{4}} \label{eq:growthRateNoFlow}
\end{align}
are the mode frequency and growth rate respectively.
Inspecting this solution, we see that the mode can be nonlinearly unstable (i.e. have a real growth rate) only if $(k_{b}^{2} - k_{c}^{2})(k_{a}^{2} - k_{b}^{2}) > 0$, which is equivalent to $k_{a} \leq k_{b} \leq k_{c}$ or $k_{c} \leq k_{b} \leq k_{a}$. Thus, to lowest order we find the traditional nonlinear coupling present in a resonant three wave decay process \cite{HasegawaThreeWaveCoupling1979}. This solution is verified against GENE simulations in figure \ref{fig:onePumpNoShear}. Figure \ref{fig:onePumpNoShear} shows that, after a short transient in the GENE simulations, there is good agreement for the mode growth rate. 
To see the impact of flow shear we must go to next order in the expansion.

\begin{figure}
  \centering
  \includegraphics[width=0.65\textwidth]{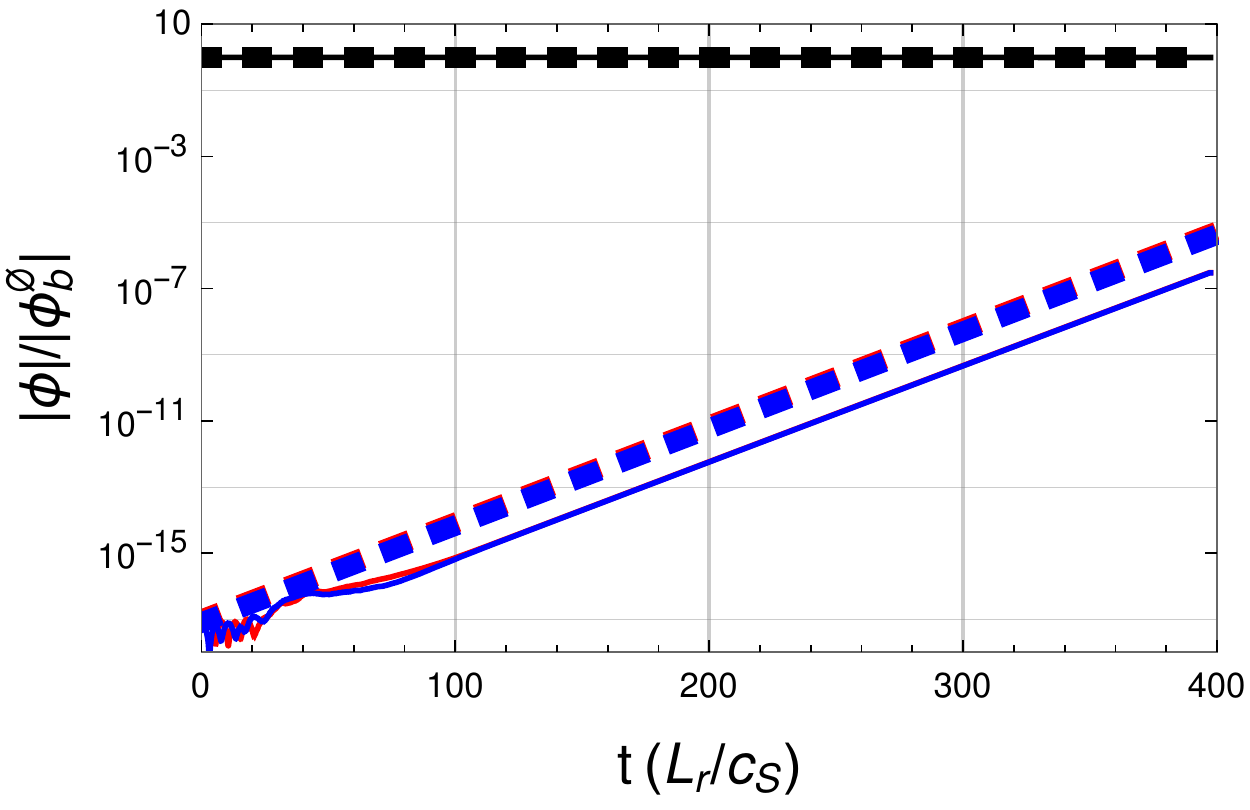}
  \includegraphics[width=0.6\textwidth]{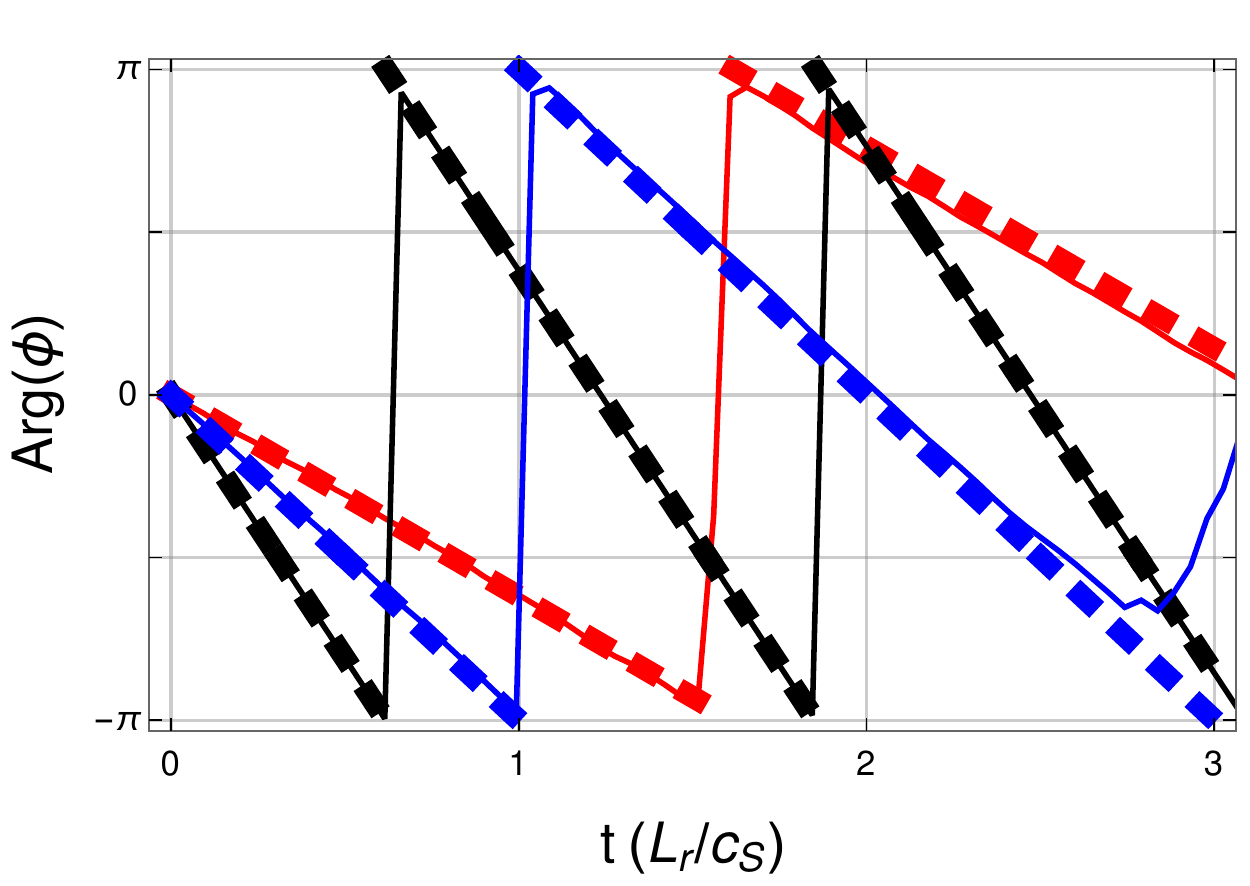}
  \caption{The amplitude (top) and phase (bottom) of the pump mode (black) and both driven modes (red and blue) as calculated by equation (\ref{eq:pumpModeSol}) (thick black dashed), equation (\ref{eq:solNoFlow}) (thick red/blue dashed), and GENE (thin solid). The parameters used were $\phi_{b}^{\varnothing}/ (B \rho_{S}^{2}) = 8.5 c_{S}/L_{r}$, $\vec{K}_{1} \rho_{S} = (-0.2,0.1)$, $\vec{K}_{2} \rho_{S} = (0.3,0.3)$, $\omega_{V\perp} = \omega_{Mx} = 0$, and $\omega_{My} = 20 c_{S}/L_{r}$.}
  \label{fig:onePumpNoShear}
\end{figure}

To next order, the differential equation for the driven mode becomes
\begin{align}
  \left. \frac{\partial^{2} \hat{\phi}_{a1}}{\partial t^{2}} \right|_{K_{x}} &- i \theta_{0} \left. \frac{\partial \hat{\phi}_{a1}}{\partial t} \right|_{K_{x}} - \left( \Lambda_{abc} \Lambda_{cab} \left| \phi_{b} \right|^{2} \right)_{0} \hat{\phi}_{a1} \label{eq:drivenModeEvolutionNextOrder} \\
  &= \left( i \theta_{1} - G_{abc1} \right) \left. \frac{\partial \hat{\phi}_{a0}}{\partial t} \right|_{K_{x}} + \left( \Lambda_{abc} \Lambda_{cab} \left| \phi_{b} \right|^{2} \right)_{1} \hat{\phi}_{a0} , \nonumber
\end{align}
where
\begin{align}
  \theta_{1} &= \omega_{V\perp} t \left( \frac{k_{ay} \rho_{S} (\omega_{Mx} - 2 K_{ax} \rho_{S} W_{a})}{1 + k_{a}^{\varnothing 2} \rho_{S}^{2}} + \frac{k_{by} \rho_{S} (\omega_{Mx} - 2 K_{bx} \rho_{S} W_{b})}{1 + k_{b}^{\varnothing 2} \rho_{S}^{2}} \right. \nonumber \\
  &\hspace{4em} \left. + \frac{k_{cy} \rho_{S} (\omega_{Mx} - 2 K_{cx} \rho_{S} W_{c})}{1 + k_{c}^{\varnothing 2} \rho_{S}^{2}} \right) \\
  G_{abc1} &= 2 \omega_{V\perp} \left( - \frac{K_{bx} k_{by} - K_{cx} k_{cy}}{k_{b}^{\varnothing 2} - k_{c}^{\varnothing 2}} + \frac{K_{bx} k_{by} \rho_{S}^{2}}{1 + k_{b}^{\varnothing 2} \rho_{S}^{2}} + \frac{K_{cx} k_{cy} \rho_{S}^{2}}{1 + k_{c}^{\varnothing 2} \rho_{S}^{2}} \right) \\
  \left( \Lambda_{abc} \Lambda_{cab} \left| \phi_{b} \right|^{2} \right)_{1} &= 2 \omega_{V\perp} t \left( \Lambda_{abc} \Lambda_{cab} \left| \phi_{b} \right|^{2} \right)_{0}  \left( \frac{K_{ax} k_{ay} - K_{bx} k_{by}}{k_{a}^{\varnothing 2} - k_{b}^{\varnothing 2}} + \frac{K_{bx} k_{by} - K_{cx} k_{cy}}{k_{b}^{\varnothing 2} - k_{c}^{\varnothing 2}} \right. \nonumber \\
  &\hspace{4em} \left. - \frac{K_{ax} k_{ay} \rho_{S}^{2}}{1 + k_{a}^{\varnothing 2} \rho_{S}^{2}} - \frac{2 K_{bx} k_{by} \rho_{S}^{2}}{1 + k_{b}^{\varnothing 2} \rho_{S}^{2}} - \frac{K_{cx} k_{cy} \rho_{S}^{2}}{1 + k_{c}^{\varnothing 2} \rho_{S}^{2}} \right) .
\end{align}
Flow shear enters through $\theta_{1}$ and $\left( \Lambda_{abc} \Lambda_{cab} \left| \phi_{b} \right|^{2} \right)_{1}$, which depend linearly on time, and $G_{abc1}$, which is constant. This equation can be solved analytically. The homogeneous solution has the same form as the lowest order solution $\hat{\phi}_{a0}$, while the particular solution is
\begin{align}
  \hat{\phi}_{a1} = \frac{1}{4 \gamma^{2}} \Big\{ C_{\iota} &\Exp{\left( i \omega + \gamma \right) t} \nonumber \\
  &\times \left[ \left( \omega - i \gamma \right) \left( \left( 1 - \gamma t \right) \theta_{1} - 2 i \gamma t G_{abc1} \right) - \left( 1 - \gamma t \right) \left( \Lambda_{abc} \Lambda_{cab} \left| \phi_{b} \right|^{2} \right)_{1} \right] \nonumber \\
  + C_{\kappa} &\Exp{\left( i \omega - \gamma \right) t} \label{eq:drivenModeSolNextOrder} \\
  &\times \left[ \left( \omega + i \gamma \right) \left( \left( 1 + \gamma t \right) \theta_{1} + 2 i \gamma t G_{abc1} \right) - \left( 1 + \gamma t \right) \left( \Lambda_{abc} \Lambda_{cab} \left| \phi_{b} \right|^{2} \right)_{1} \right] \Big\} . \nonumber
\end{align}
Thus, we see that the dominant effect of weak flow shear (or alternatively the first effect of flow shear to appear) is a quadratic correction to the lowest order exponential behavior. This is similar to the effect of flow shear on the PVG-driven instability that we found in section \ref{subsec:PVGlowFlowLimit}.

However, we notice that this solution, which is fairly complex, can be simplified. Specifically, we are primarily interested in the behavior after a few e-folding times (i.e. $\gamma t \gg 1$ while retaining $|\omega_{V\perp} t| \ll 1$), rather than the details of the transients at the very beginning of the evolution. This limit is equivalent to $\omega_{V\perp} \ll \gamma$ and the dominant terms are those that are quadratic in time, giving
\begin{align}
  \hat{\phi}_{a1} = \frac{t}{4 \gamma} \Big[ C_{\iota} &\left[ \left( \Lambda_{abc} \Lambda_{cab} \left| \phi_{b} \right|^{2} \right)_{1} - \left( \omega - i \gamma \right) \theta_{1} \right] \Exp{\left( i \omega + \gamma \right) t} \nonumber \\
  - C_{\kappa} &\left[ \left( \Lambda_{abc} \Lambda_{cab} \left| \phi_{b} \right|^{2} \right)_{1} - \left( \omega + i \gamma \right) \theta_{1} \right] \Exp{\left( i \omega - \gamma \right) t} \Big] .
\end{align}
As in section \ref{subsec:PVGlowFlowLimit}, for unstable modes (i.e. $\gamma$ is real), the $\Exp{\left( i \omega + \gamma \right) t}$ term will dominate and the real part of its coefficient indicates if flow shear will enhance or stabilize its growth. Therefore, the instability will be enhanced by perpendicular flow shear if and only if $\left( \Lambda_{abc} \Lambda_{cab} \left| \phi_{b} \right|^{2} \right)_{1} > \theta_{0} \theta_{1} / 2$. In other words, the instability is enhanced if the finite flow shear correction to the growth rate (i.e. equation \refEq{eq:growthRateNoFlow}) is positive. This result should be expected according to our previous discussions of the $\omega_{V\perp} \ll \gamma$ limit. If flow shear is moving the mode to a wavenumber that is more strongly driven (relative to damping), then it should begin to grow faster.

\subsubsection{The $|\omega_{V\perp} t| \gg 1$ limit}

\begin{figure}
  \centering
  \includegraphics[width=0.65\textwidth]{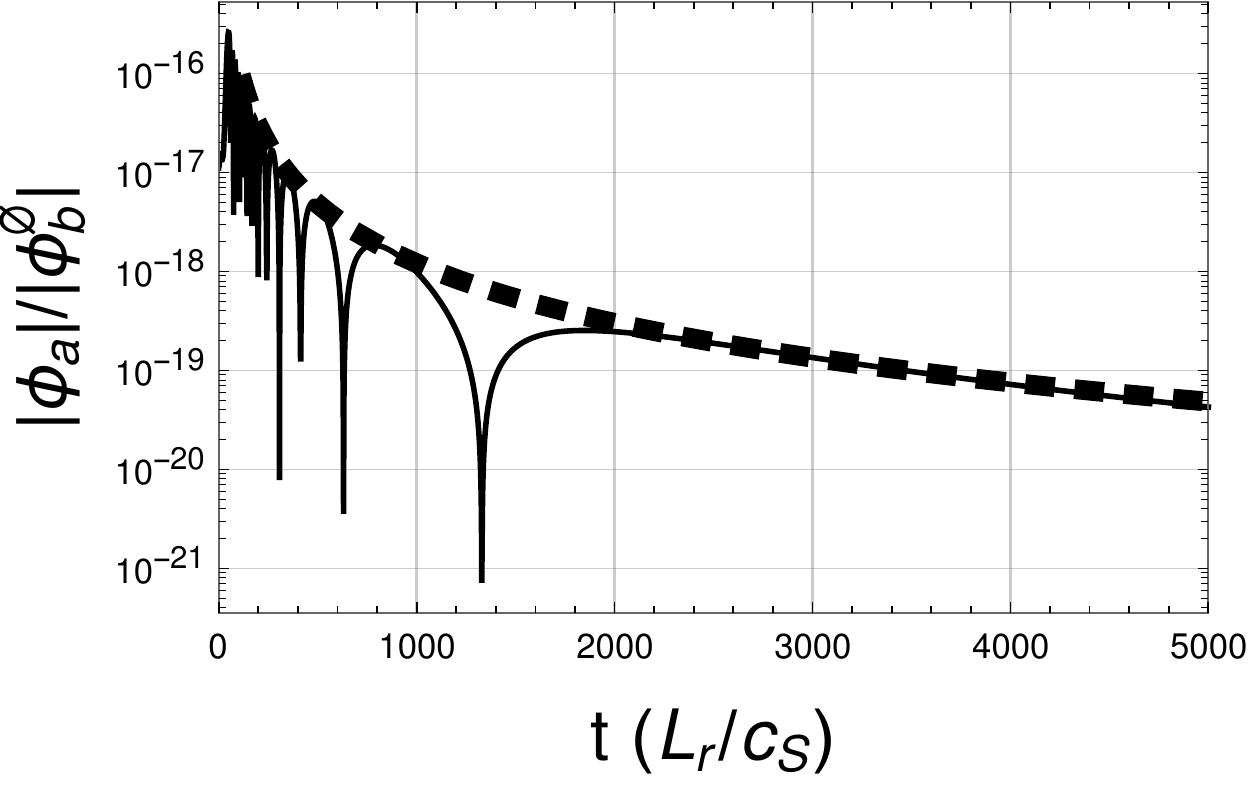}
  \caption{The amplitude of the $\vec{K}_{a} \rho_{S} = (-0.15,0.2)$ driven mode as calculated by a numerical solution to equation (\ref{eq:drivenModeEvolution}) (solid) and the long time $t^{-2}$ scaling predicted by equation \refEq{eq:longTimeSolNonZonal} (dashed). The parameters used were identical to figure \ref{fig:onePumpDrivenShear} except $\omega_{V\perp} = -0.1 c_{S}/L_{r}$.}
  \label{fig:onePumpDrivenShearLongTime}
\end{figure}

To lowest order in the long time limit, the coefficients in equation \refEq{eq:drivenModeEvolution} reduce to
\begin{align}
  \theta_{0} ~ t =& ~ \frac{\omega_{Mx}}{\omega_{V\perp}} \left( \frac{1}{k_{ay} \rho_{S}} + \frac{1}{k_{by} \rho_{S}} + \frac{1}{k_{cy} \rho_{S}} \right) ~~ \sim \Order{1} \\
  G_{abc0} ~ t =& ~ 2 ~~ \sim \Order{1} \\
  \left( \Lambda_{abc} \Lambda_{cab} \left| \phi_{b} \right|^{2} \right)_{0} t^{2} =& ~ \frac{1}{B^{2}} \left( \vec{K}_{b} \times \vec{K}_{c} \right) \cdot \hat{b} \left( \vec{K}_{a} \times \vec{K}_{b} \right) \cdot \hat{b} \\ &\times \frac{k_{by}^{2} - k_{cy}^{2}}{k_{ay}^{2}} \frac{k_{ay}^{2} - k_{by}^{2}}{k_{cy}^{2}} \frac{\left( 1 + k_{b}^{\varnothing 2} \rho_{S}^{2} \right)^{2}}{k_{by}^{4} \rho_{S}^{4} \omega_{V\perp}^{4} t^{2}} \left| \phi_{b}^{\varnothing} \right|^{2} ~~ \sim \Order{\omega_{V\perp}^{-2} t^{-2}} . \nonumber
\end{align}
We see that the nonlinear coupling term is small, which means that we can return to equations \refEq{eq:mode1evolution} and \refEq{eq:mode3evolution} and solve them on their own. As should be expected the nonlinear coupling term
\begin{align}
  \left( \Lambda_{abc} \hat{\phi}_{b}^{\ast} \frac{1 + k_{a}^{2} \rho_{S}^{2}}{(1 + k_{b}^{2} \rho_{S}^{2}) (1 + k_{c}^{2} \rho_{S}^{2})} \right)_{0} t =& - \frac{1}{B} \left( \vec{K}_{b} \times \vec{K}_{c} \right) \cdot \hat{b} \\
  &\times \frac{k_{by}^{2} - k_{cy}^{2}}{k_{ay}^{2}} \frac{k_{ay}^{2}}{k_{by}^{2} k_{cy}^{2} \rho_{S}^{2} \omega_{V\perp}^{2} t} \hat{\phi}_{b}^{\ast} ~~ \sim \Order{\omega_{V\perp}^{-1} t^{-1}} \nonumber
\end{align}
is small there too, so that the lowest order solution is simply
\begin{align}
  \hat{\phi}_{a0} &= C_{\lambda} \label{eq:longTimeSolNonZonal} \\
  \hat{\phi}_{c0} &= C_{\mu} . \label{eq:longTimeSolNonZonalModeC}
\end{align}
This limit is verified against the full numerical solution in figure \ref{fig:onePumpDrivenShearLongTime}. This means that the driven modes are not driven at all and simply evolve in isolation according to the $t^{-2}$ dependence contained in equation \refEq{eq:hatDef}. Thus, they either decay as $t^{-2}$ as they move towards $k_{x} \rightarrow \pm \infty$ or grow as $t^{2}$ as flow shear brings them towards $k_{x} = 0$ from large $|k_{x}|$. To next order, the coupling term enters equations \refEq{eq:mode1evolution} and \refEq{eq:mode3evolution}, which can be solved to find
\begin{align}
  \hat{\phi}_{a1} &= - \left( \Lambda_{abc} \hat{\phi}^{\ast}_{b} \frac{1 + k_{a}^{2} \rho_{S}^{2}}{(1 + k_{b}^{2} \rho_{S}^{2}) (1 + k_{c}^{2} \rho_{S}^{2})} \right)_{0} t ~ \hat{\phi}^{\ast}_{c0} \frac{1 + i \theta_{0} t}{1 + (\theta_{0} t)^{2}} \left(\frac{t}{t^{\varnothing}}\right)^{i \theta_{0} t} , \label{eq:longTimeSolNonZonalNextOrder}
\end{align}
where $t^{\varnothing}$ is the time at which $C_{\lambda}$ and $C_{\mu}$ are determined. The expression for $\hat{\phi}_{c1}$ is the same except the ``a'' and ``c'' indexes must be swapped.

From equations \refEq{eq:longTimeSolNonZonal} and \refEq{eq:longTimeSolNonZonalModeC}, we see that the dominant effect of flow shear at long times is the $t^{-2}$ dependence from the ion polarization drift factor in equation \refEq{eq:hatDef}. The dominant effect of the nonlinear coupling on the amplitude of a decaying mode is given by equations \refEq{eq:hatDef} and \refEq{eq:longTimeSolNonZonalNextOrder}, which scales as $t^{-3}$.

\subsubsection{Zonal pump modes}

When the pump mode is zonal, the solution in the $|\omega_{V\perp} t| \ll 1$ limit remains largely the same. However, the behavior in the $|\omega_{V\perp} t| \gg 1$ limit does not. In contrast to a non-zonal pump mode, the nonlinear coupling is {\it not} small for a zonal pump mode. Even at long times the nonlinear interaction is still important. To lowest and next order, the coefficients in equation \refEq{eq:drivenModeEvolution} are
\begin{align}
  \theta_{0} t + \theta_{1} t =& ~ \frac{K_{bx} \rho_{S} \omega_{Mx} t}{1 + K_{bx}^{2} \rho_{S}^{2}} - \frac{\left( K_{ax} + K_{cx} \right) \omega_{Mx}}{k_{a y}^{2} \rho_{S} \omega_{V\perp}^{2} t} ~~ \sim \Order{\omega_{V\perp} t} \\
  G_{abc0} t + G_{abc1} t =& ~ 0 ~~ \sim \Order{\omega_{V\perp}^{-2} t^{-2}} \\ 
  \left( \Lambda_{abc} \Lambda_{cab} \left| \phi_{b} \right|^{2} \right)_{0} t^{2} +& \left( \Lambda_{abc} \Lambda_{cab} \left| \phi_{b} \right|^{2} \right)_{1} t^{2} \label{eq:zonalLongTimeCoupling}  \\
  =& ~ - \frac{K_{bx}^{2} k_{ay}^{2}}{B^{2}} \left| \phi_{b}^{\varnothing} \right|^{2} t^{2} + 2 \frac{K_{bx}^{2} \left( 1 + K_{bx}^{2} \rho_{S}^{2} \right)}{B^{2} \rho_{S}^{2} \omega_{V\perp}^{2}} \left| \phi_{b}^{\varnothing} \right|^{2} ~~ \sim \Order{\omega_{V\perp}^{2} t^{2}} \nonumber
\end{align}
and for the equation to balance we require that $\partial/\partial t \sim \omega_{V\perp}$. This means that once again we must solve equation \refEq{eq:drivenModeEvolution} with constant coefficients (and $G_{abc} = 0$). Thus, like in section \ref{sec:shortTimeNonlinear}, the lowest order solution is given by
\begin{align}
  \hat{\phi}_{a0} = C_{\nu} \Exp{ i \left( \omega + \gamma \right) t} + C_{\xi} \Exp{ i \left( \omega - \gamma \right) t} , \label{eq:zonalLongTimeSol}
\end{align}
where
\begin{align}
  \omega &\equiv \frac{\theta_{0}}{2} \\
  \gamma &\equiv \sqrt{- \left( \Lambda_{abc} \Lambda_{cab} \left| \phi_{b} \right|^{2} \right)_{0} + \frac{\theta_{0}^{2}}{4}} .
\end{align}
The important distinction, which we have made explicit in equation \refEq{eq:zonalLongTimeSol}, is that the lowest order nonlinear coupling term is necessarily negative. Thus, the nonlinear coupling makes the mode oscillate and modifies the phase of the mode, which can have important consequences on which other modes it can resonate with. However, it does not affect the scaling of the mode amplitude. To lowest order, the dominant effect of flow shear is the $t^{-2}$ factor from the ion polarization drift in equation \refEq{eq:hatDef}. This scaling is verified in figure \ref{fig:onePumpDrivenShearLongTimeZonal} for a mode that is decaying in amplitude as flow shear advects it to $k_{x} \rightarrow \infty$. The oscillation frequency predicted by equation \refEq{eq:zonalLongTimeSol} was also found to agree.

\begin{figure}
  \centering
  \includegraphics[width=0.65\textwidth]{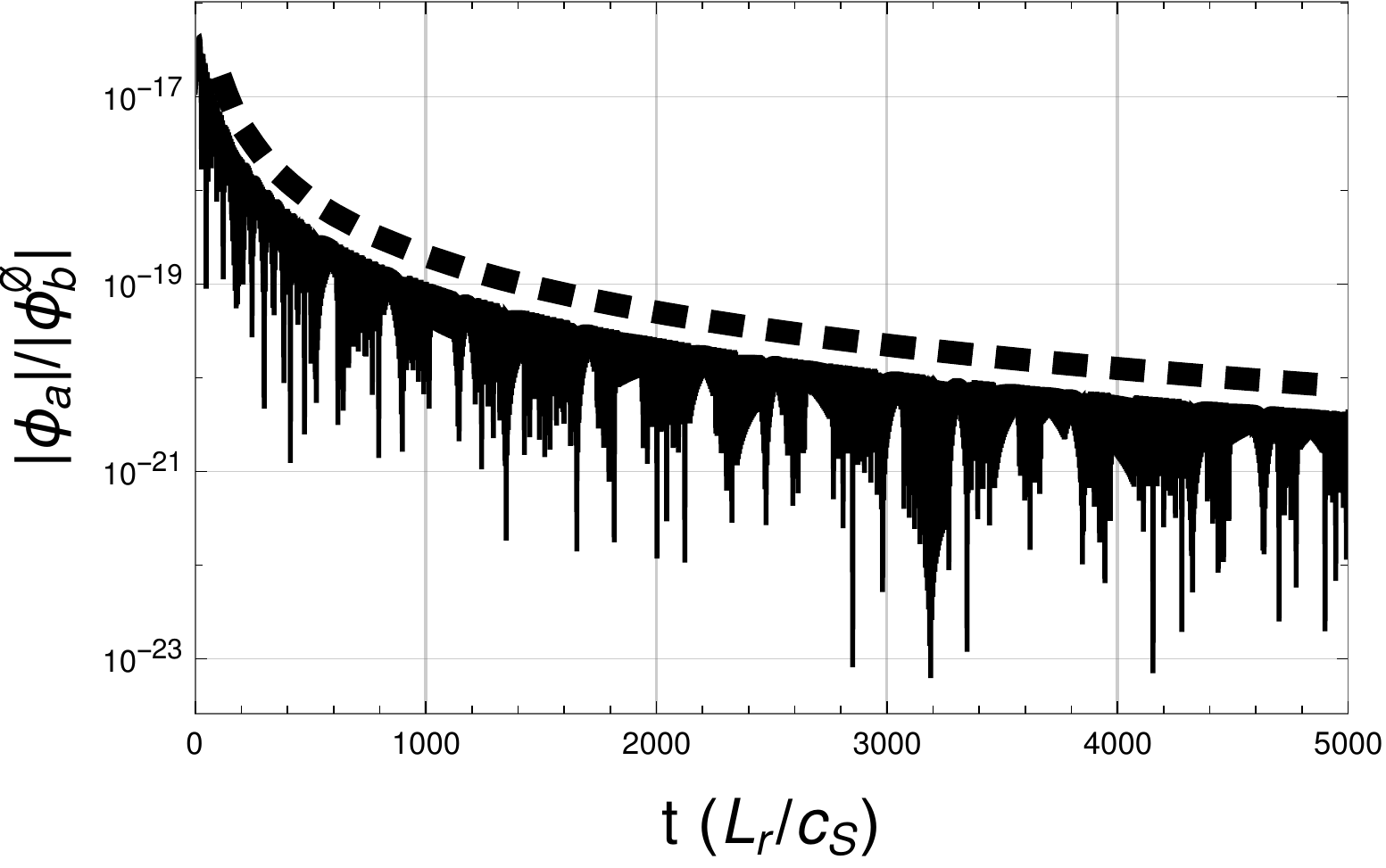}
  \caption{The amplitude of the $\vec{K}_{a} \rho_{S} = (-0.15,0.2)$ driven mode as calculated by a numerical solution to equation (\ref{eq:drivenModeEvolution}) (solid) and the long time $t^{-2}$ scaling predicted by equations \refEq{eq:hatDef} and \refEq{eq:zonalLongTimeSol} (dashed). The parameters used were identical to figure \ref{fig:onePumpDrivenShear} except $\omega_{V\perp} = -0.1 c_{S}/L_{r}$ and $\vec{K}_{b} \rho_{S} = (0.4,0)$.}
  \label{fig:onePumpDrivenShearLongTimeZonal}
\end{figure}

To the next nontrivial order, $\Order{\omega_{V\perp}^{2} t^{2}}$, the differential equation gains time dependent coefficients. It becomes identical to equation \refEq{eq:drivenModeEvolutionNextOrder}, except the first order coefficients are proportional to $t^{-2}$ instead of $t$ (and $G_{abc1} = 0$). The inhomogeneous solution to this is 
\begin{align}
  \hat{\phi}_{a1} = C_{\nu} \Exp{i \left( \omega - \gamma \right) t} & E_{I} \left( 2 i \gamma t \right) \left[ \left( \omega + \gamma \right) \theta_{1} - \left( \Lambda_{abc} \Lambda_{cab} \left| \phi_{b} \right|^{2} \right)_{1} \right] \\
  + C_{\xi} \Exp{i \left( \omega + \gamma \right) t} & E_{I} \left( - 2 i \gamma t \right) \left[ \left( \omega - \gamma \right) \theta_{1} - \left( \Lambda_{abc} \Lambda_{cab} \left| \phi_{b} \right|^{2} \right)_{1} \right] , \nonumber
\end{align}
where $E_{I} \left( z \right) \equiv - \int_{-z}^{\infty} dt \Exp{-t}/t$ is the exponential integral function. As before, we can expand in $\gamma t \gg 1$ to get the behavior after a few e-folding times. We find
\begin{align}
  \hat{\phi}_{a1} = - C_{\nu} \Exp{i \left( \omega + \gamma \right) t} & \frac{i}{2 \gamma t} \left[ \left( \omega + \gamma \right) \theta_{1} - \left( \Lambda_{abc} \Lambda_{cab} \left| \phi_{b} \right|^{2} \right)_{1} \right] \\
  + C_{\xi} \Exp{i \left( \omega - \gamma \right) t} & \frac{i}{2 \gamma t} \left[ \left( \omega - \gamma \right) \theta_{1} - \left( \Lambda_{abc} \Lambda_{cab} \left| \phi_{b} \right|^{2} \right)_{1} \right] , \nonumber
\end{align}
which has a purely imaginary exponent and is proportional to $t^{-3}$. Thus, once again, our solution does not affect the amplitude of the decaying mode. This means that, factoring in the $t^{-2}$ dependence from equation \refEq{eq:hatDef}, the effect of the nonlinear coupling term on the amplitude of a driven mode decays away more rapidly than $t^{-5}$.

Comparing to the previous subsection, we find a surprising result. The long time effect of the nonlinear coupling on the mode amplitude is much weaker when the pump mode is zonal. Specifically, the coupling from a non-zonal pump scales as $t^{-3}$, while the coupling from a zonal pump scales more weakly than $t^{-5}$. This suggests that coupling between non-zonal modes has the dominant nonlinear effect on the amplitude of a mode when it is at large $k_{x}$. Intuitively, it seems like the opposite should be true because the zonal pump does not vary much with time, while a non-zonal pump does. Nevertheless, the coupling with zonal modes still has the dominant effect on the phase and oscillation frequency of a mode and this effect does not get weaker when the mode is at high $k_{x}$.

\section{Conclusions}
\label{sec:conclusions}

In this work we have taken the cold ion limit of local $\delta f$ gyrokinetics to derive an exact yet simple fluid model (i.e. equations \refEq{eq:phiEvolution} and \refEq{eq:velocityMoment}). This model was applied to a slab geometry that included magnetic drifts, but no magnetic shear. The resulting equations were found to reduce to the Charney-Hasegawa-Mima model in the presence of background flow. Then, an analytic solution (i.e. equation \refEq{eq:phiLinearSol}) was found for the full time evolution of a single Fourier mode driven unstable by a parallel velocity gradient, but stabilized by perpendicular flow shear. Studying this solution revealed that quite complicated behavior is possible (e.g. figure \ref{fig:stableButUnbounded}). Additionally, we calculated simple criteria governing the initial stability (i.e. equation \refEq{eq:PVGstabilityShortTimekPar}) and long time stability (i.e. equation \refEq{eq:unboundedPVGcondSimple}) of a mode. These two criteria were found to be somewhat different.

The fluid model was also used to study the basic three-mode nonlinear coupling in two-dimensional turbulence. The initial nonlinear drive of zonal flows was investigated using two large pump modes (i.e. equation \refEq{eq:mode1SolTwoPump}) and the analytic results were verified against nonlinear gyrokinetic simulations (i.e. figures \ref{fig:twoPumpPump} and \ref{fig:twoPumpDriven}). Next, the physics occurring in the quasi-steady-state of nonlinearly saturated turbulence was studied using one large pump mode. Specifically, we focused on how a single large zonal mode drives and regulates non-zonal modes. We found that coupling with a zonal mode has an important effect on the {\it phase and oscillation frequency} of a mode that does not get weaker as flow shear advects the mode from high $k_{x}$ or to high $k_{x}$. However, coupling with non-zonal modes actually has the dominant effect on the {\it amplitude} of a mode. Again, the results were verified against gyrokinetic simulations (i.e. figures \ref{fig:onePumpPumpShear} through \ref{fig:onePumpDrivenShearLongTimeZonal}).

\section*{Acknowledgments}

This work has been carried out within the framework of the EUROfusion Consortium and has received funding from the Euratom research and training programme 2014-2018 under grant agreement No 633053. The views and opinions expressed herein do not necessarily reflect those of the European Commission. Simulations were performed with the support of EUROfusion and MARCONI-Fusion. This work was supported by a grant from the Swiss National Supercomputing Centre (CSCS) under project ID s863. This work was supported in part by the Swiss National Science Foundation.

\section*{References}
\bibliographystyle{unsrt}
\bibliography{references.bib}

\end{document}